\documentclass{article}

\usepackage{amsmath,amssymb}
\usepackage{graphicx}
\usepackage{color,xcolor}

\usepackage{hyperref}
\usepackage{cite}
\usepackage{mathrsfs}
\usepackage{mathbbol}
\usepackage{amsfonts}
\usepackage{amsthm}
\usepackage{url}

\newtheorem{fact}{Fact}

\newtheorem{theorem}{Theorem}
\newtheorem{lemma}{Lemma}
\newtheorem{corollary}{Corollary}

\newtheorem{definition}{Definition}

\newtheorem{remark}{Remark}

\newtheorem{condition}{Condition} %
\newtheorem{statement}{Statement} %
\newtheorem{example}{Example}%

\newcommand{\pf}{\begin{proof}}  
\newcommand{\fp}{\end{proof}} 
\newcommand{\set}[1]{\left\{#1\right\}} 
\newcommand{\cset}[1]{\left|#1\right|} 
\newcommand{\comp}{\circ} 

\newcommand{\Int}{\mathbb{Z}} 
\newcommand{\GF}[1]{\mathbb{F}_{#1}} 
\newcommand{\uGF}[1]{\mathbb{F}_{#1}^{*}} 
\newcommand{\unit}[1]{#1^*}

\newcommand{\seq}[1]{\underline{#1}}

\newcommand{\red}[1]{\overline{#1}} 
\newcommand{\indmap}[1]{\widehat{#1}} 
\newcommand{\pring}[1][p^e]{\Int/{#1}\Int} 

\newcommand{\rlt}[1]{{#1}} 
\newcommand{\rlab}[1][e]{{R}_{#1}\left(\seq{u},\seq{v}\right)} 
\newcommand{\rlabs}[1][e]{{R}_{#1}^{\mathrm{S}}\left(\seq{u},\seq{v}\right)} 
\newcommand{\rlabe}[1][e]{{R}_{#1}^{\mathrm{E}}\left(\seq{u},\seq{v}\right)} 
\newcommand{\brlab}[1][e]{\widetilde{{R}}_{#1}\left(\seq{u},\seq{v}\right)} 
\newcommand{\pair}[2]{\left(#1,#2\right)}

\newcommand{\gseqg}[1][]{G_{#1}(\prmpoly,p^e)} 
\newcommand{\gseq}[1][]{G'_{#1}(\prmpoly,p^e)} 

\newcommand{\prmpoly}{\sigma(x)} 

\newcommand{\len}{n} 
\newcommand{\rep}[1]{#1} 
\newcommand{\crd}[2][i]{\left\langle#2\right\rangle_{#1}} 
\newcommand{\crr}[3][\omega]{\left\{#3\right\}_{#1,#2}} 
\newcommand{\cut}[2][e-1]{\left[#2\right]_{#1}} 

\newcommand{\tail}{\epsilon} 

\newcommand{\gset}{\unit{(\pring)}} 
\newcommand{\ggset}{A} 

\newcommand{\perb}{\left(p^\len-1\right)}
\newcommand{\cmap}{\varphi} 
\newcommand{\lindep}{\sim}
\newcommand{\linind}{\not\sim}
\newcommand{\imset}{S}

\renewcommand{\delta}{h(x)}

\newcommand{\ls}[1]
    {\dimen0=\fontdimen6\the\font\lineskip=#1\dimen0
     \advance\lineskip.5\fontdimen5\the\font
     \advance\lineskip-\dimen0
     \lineskiplimit=0.9\lineskip
     \baselineskip=\lineskip
     \advance\baselineskip\dimen0
     \normallineskip\lineskip\normallineskiplimit\lineskiplimit
     \normalbaselineskip\baselineskip
     \ignorespaces}

\begin{document}

\bibliographystyle{abbrv}

\title{Injectivity w.r.t. Distribution of Elements in the Compressed Sequences
Derived from Primitive Sequences over $Z/p^eZ$}
\author{Lin Wang\\
Science and Technology on Communication Security Laboratory\\
Chengdu, 610041, P.R. China\\
Email. linwang@math.pku.edu.cn\\
        Zhi Hu\\
School of Mathematics and Statistics, Central South University\\
 Changsha, 410083 Hunan, P.R. China\\
huzhi\_math@csu.edu.cn\\
}

\date{}
 \maketitle

\thispagestyle{plain}
\setcounter{page}{1}

\begin{abstract}
Let $p\geq3$ be a prime and  $e\geq2$ an integer.
Let $\prmpoly$ be a primitive polynomial of degree $\len$ over $\pring$
and $\gseq$ the set of primitive linear recurring sequences generated by $\prmpoly$.
A compressing map $\cmap$ on $\pring$ naturally induces
a map $\indmap{\cmap}$ on $\gseq$.
For a subset $D$ of the image of $\cmap$,
$\indmap{\cmap}$ is called to be injective w.r.t. $D$-uniformity
if the distribution of elements of $D$ in the compressed sequence
implies all information of the original primitive sequence.
In this correspondence, for at least $1-2(p-1)/(p^\len-1)$ of
primitive polynomials of degree $\len$, a clear criterion on $\cmap$
is obtained to decide whether $\indmap{\cmap}$ is injective w.r.t. $D$-uniformity,
and the majority of maps on $\pring$ induce injective maps on $\gseq$.
Furthermore, a sufficient condition on $\cmap$ is given
to ensure injectivity of $\indmap{\cmap}$ w.r.t. $D$-uniformity.
It follows from the sufficient condition that
if $\prmpoly$ is strongly primitive and
the compressing map $\cmap(x)=f(x_{e-1})$, where $f(x_{e-1})$
is a permutation polynomial over $\GF{p}$,
then $\indmap{\cmap}$ is injective w.r.t. $D$-uniformity
for $\emptyset\neq D\subset\GF{p}$.
Moreover, we give three specific families of compressing maps
which induce injective maps on $\gseq$.\\


\end{abstract}

\ls{1.5}
\section{Introduction}\label{sect:intro}
%
Pseudorandom sequences play a significant role in
 coding, cryptography and digital communication systems.
A linear feedback shift register (LFSR) over a residue ring $\pring$
is a candidate to
construct pseudorandom generators. For example, the stream cipher
ZUC \cite{ZUC} for \textquotedblleft4G\textquotedblright\, mobile
standard Long Term Evolution employs an LFSR over the residue ring
$\Int/(2^{31}-1)\Int$. As a result of an extended Berlekamp-Massey
algorithm by Reeds and Sloane\cite{RS85}, linear recurring sequences
generated by an LFSR over residue rings can be synthesized
efficiently. Hence, in cryptographic scenarios compressing maps are
used to derive nonlinear sequences from linear recurring sequences
over residue rings
\cite{dai92,huang88,kzm2,KN95,KN93,nchv91dm,qwf01,qwf98,ward,ZDH}, and such
compressed sequences were proposed as candidates for the keystream
of a stream cipher\cite{SQ,ZQ04,ZQ07it}. It was also shown that
certain compressed sequences meet some pseudorandom properties,
e.g., distribution of zeros and ones, autocorrelation and linear
complexity\cite{DBG,FH,kzm2,QZh1,QZh2,SZ}.

\begin{figure}[htbn]
\begin{center}
\setlength{\unitlength}{1mm}
\begin{picture}(50,16)
\put(5,0){\framebox(30,6){Ring-LFSR}} \put(0,3){\vector(1,0){5}}

\put(10,6){\line(0,1){2.5}} \put(10,10){\oval(12,3)}
\put(6,9.5){$-c_{n-1}$} \put(10,11.5){\line(0,1){2.5}}
\put(8.75,14){$\oplus$}

\put(30,6){\line(0,1){2.5}} \put(30,10){\oval(12,3)}
\put(28,9.5){$-c_{0}$} \put(30,11.5){\line(0,1){3.5}}

\put(30,15){\line(-1,0){5}} \put(15,15){\line(-1,0){4}}
\put(18,14){$\cdots$} \put(18,9.5){$\cdots$}

\put(9,15){\line(-1,0){9}} \put(0,15){\line(0,-1){12}}

\put(35,3){\vector(1,0){3}} \put(40,3){\circle{4}}
\put(39,2){$\cmap$} \put(42,3){\vector(1,0){3}} \put(45,3){output}
\end{picture}
\end{center}
\caption{A 
PRNG by ring-LFSR}\label{fig:RLFSR}
\end{figure}
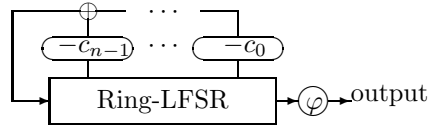

As in Fig.\ref{fig:RLFSR}, such a sequence generator consists of a
compressing map $\cmap$ defined on $\pring$ and an LFSR
described by its characteristic polynomial
$\prmpoly=x^\len+c_{\len-1}x^{\len-1}+\cdots+c_0$, $c_i\in \pring$.
Specifically,
a linear recurring sequence $\seq{s}$ generated by this LFSR 
abides by 
\begin{equation}\label{eqn:lfsr}
    \seq{s}(t) = -c_{\len-1}\seq{s}(t-1)-\cdots-c_0\seq{s}(t-\len).
\end{equation}
If $\min\set{m>0:\prmpoly\mid x^m-1}=p^{e-1}(p^\len-1)$, then
$\prmpoly$ is said to be \emph{primitive}. For $p\geq3$, let $h(x)$
be a polynomial over $\GF{p}$ satisfying $h(x)\equiv
\left(x^{p^\len-1}-1\right)/p \mod (p,\prmpoly)$ and $\deg h(x) <
\deg\prmpoly$. If $\prmpoly$
is primitive and 
$\deg h(x)\geq1$, then $\prmpoly$ is said to be \emph{strongly
primitive}.
A maximal length sequence generated by $\prmpoly$
is called to be \emph{primitive}.

Let $\gseq$ denote the set of primitive 
sequences generated by $\prmpoly$.
The compressing map $\cmap$ in Fig.\ref{fig:RLFSR} naturally induces
a map $\indmap{\cmap}$ on $\gseq$.
Specifically, $\indmap{\cmap}$ maps a sequence
$\seq{s}=\left(\dots, \seq{s}(t-1), \seq{s}(t), \seq{s}(t+1), \dots\right)$ to
$\indmap{\cmap}(\seq{s})=\left(\dots, \cmap(\seq{s}(t-1)), \cmap(\seq{s}(t)), \cmap(\seq{s}(t+1)), \dots\right)$.
Here $\indmap{\cmap}(\seq{s})$ is called the compressed sequence of $\seq{s}$.
A research problem is to decide whether
partial (or all)  information of the compressed sequence
identifies a unique original primitive sequence.
Many papers studied this problem from the following two aspects.

One aspect is the injectivity of the induced map $\indmap{\cmap}$.
If $\indmap{\cmap}$ is injective on $\gseq$, then the compressing map
$\cmap$ is said to be \emph{entropy-preserving}\cite{qwf01,SQ}.
Entropy-preservation of
compressing maps has hitherto attracted extensive research
\cite{huang88,HD,qwf01,qwf98,SQ,TQ07,ZDH,ZQ04,ZQ07it,ZQ08}.
Zhu and Qi \cite{ZQ08} proved that the modular function
$\cmap(x)=x\bmod M$
is entropy-preserving, where 
$M\geq2$ is not a power of $p$.
Since $x\in\Int/p^e\Int$ can be identified as $x=x_0 + x_1p+\cdots +
x_{e-1}p^{e-1}$, $x_i\in\set{0,1,\dots,p-1}$, a function
$\Int/p^e\Int\rightarrow\GF{p}$ is naturally interpreted as an
$e$-variable function over $\GF{p}$, and a sequence $\seq{s}$ over
$\Int/p^e\Int$ was written uniquely as $\seq{s}=\seq{s}_0 +
\seq{s}_1p+\cdots+\seq{s}_{e-1}p^{e-1}$ in
\cite{huang88,HD,qwf01,qwf98,SQ,TQ07,ZDH,ZQ04,ZQ07it,ZQ08}, where
$\seq{s}_i$ is a sequence over $\GF{p}$.
It is known that $\cmap(x) = x_{e-1}$ is entropy-preserving,
i.e., the \emph{highest level sequence} $\seq{s}_{e-1}$
contains the same information as $\seq{s}$ \cite{huang88,HD,KN93}.
For $p\geq5$ and $e\geq2$, Zhu and
Qi \cite{ZQ04} designed a kind of  entropy-preserving maps
\begin{equation}\label{eqn:map-zq3}
\cmap(x) = x_{e-1}^\ell+f_2(x_0,\dots,x_{e-2}),
\end{equation}
where $2 \leq \ell<p$. Besides, for $p\geq3$ and $e\geq3$, Sun and
Qi \cite{SQ} found another kind of entropy-preserving maps
\begin{equation}\label{eqn:map-zq2}
\cmap(x)=x_{e-1}(g_0(x_{e-2})+g_1(x_0,x_1,\dots,x_{e-3}))+f_2(x_0,\dots,x_{e-2}),
\end{equation}
where $\deg g_0\geq2$ and $\gcd(p-1,\deg g_0+1)=1$. Furthermore,
thanks to \cite{TQ07,ZQ04,ZQ07it}, if $\prmpoly$ is
strongly primitive, then the compressing map
\begin{equation}\label{eqn:map-zq1}
\cmap(x)=f_0(x_{e-1}) + f_1(x_0,\dots,x_{e-2})
\end{equation}
is entropy-preserving, where $1\leq\deg f_0 < p$.
Recently, the papers \cite{CQ09,HW,ZxQ11,ZQT13,ZxQ13}
considered entropy-preservation of
maximal length sequences over $\Int/N\Int$,
where $N$ is an odd square-free integer.

The other aspect is whether the distribution of an element in
the compressed sequence determines a unique original sequence.
Here let $\cmap:\pring\rightarrow\GF{p}$ be the compressing map.
For $\seq{a},\seq{b}\in\gseq$ and $s,k\in\GF{p}$,
the compressed sequences
$\indmap{\cmap}(\seq{a})$ and $\indmap{\cmap}(\seq{b})$ are
called $s$-uniform, $s$-uniform with $\seq{\alpha}$,
and $s$-uniform with $\seq{\alpha}|_k$, respectively,
if $\indmap{\cmap}(\seq{a})(t)=s$ iff $\indmap{\cmap}(\seq{b})(t)=s$
for all $t\in\Int$, for all $t\in\Int$ with $\seq{\alpha}(t)\neq0$,
and for all $t\in\Int$ with $\seq{\alpha}(t)=k$,
where the sequence $\seq{\alpha}=\delta\seq{a}\bmod p$
and $\delta$ is an operator on sequences defined in Section \ref{sect:main}.
Zhu and Qi \cite{ZQ05,ZQ07} proved that $\seq{a}=\seq{b}$ iff $\indmap{\cmap}(\seq{a})$
and $\indmap{\cmap}(\seq{b})$ are $0$-uniform, i.e.,
under the compressing map $x\mapsto x_{e-1}$
the distribution of $0$ in the compressed sequence
implies all information of the original sequence.
Zheng and Qi \cite{ZxQ10,ZxQ13uni} proved that
for $s\in\GF{p}$ and $k\in\uGF{p}$,
under the compressing map $x\mapsto x_{e-1}+\eta(x_0,\dots,x_{e-2})$
with the coefficient of monomial $x_{e-2}^{p-1}\cdots x_1^{p-1}x_0^{p-1}$ not
equal to $(-1)^e(p+1)/2$,
the distribution of $s$ in the compressed sequence
at clock cycles $t$ with $\seq{\alpha}(t)=k$
implies all information of the original sequence.
Jiang and Lin \cite{JL14,JL14p} also gave fruitful results
about uniformity with $\seq{\alpha}|_k$.

\emph{Our contribution.} Let $D$ be a subset of the image set
of the compressing map $\cmap$.
We define $D$-uniformity to describe the distribution of
elements of $D$ in the compressed sequences. $D$-uniformity
generalizes the definition of $s$-uniformity and entropy-preservation.
We call the induced map $\indmap{\cmap}$ to be injective w.r.t. $D$-uniformity
if the distribution of elements of $D$ in the compressed sequence
contains all information of the original primitive sequence.
Let $p$ be an odd prime.
We find that at least $1-2(p-1)/(p^\len-1)$ of $\len$-degree
primitive polynomials $\prmpoly$ satisfy
$\left(x^{p^\len-1}-1\right)^2/p^2\not\equiv a \bmod (p,\prmpoly)$
for any $a\in\GF{p}$.
For primitive sequences generated by those polynomials,
we give a clear criterion on $\cmap$ to decide
whether $\indmap{\cmap}$ is injective w.r.t. $D$-uniformity.
Particularly, for those primitive polynomials
 our quantitative estimation suggests that
the majority of maps on $\pring$ 
are entropy-preserving.
Furthermore, we also give a sufficient condition on $\cmap$
to ensure injectivity of $\indmap{\cmap}$ w.r.t. $D$-uniformity.
Based on our sufficient condition, we show that
if $\prmpoly$ is strongly primitive and
the compressing map $\cmap(x)=f(x_{e-1})$, where $f(x_{e-1})$
is a permutation polynomial over $\GF{p}$,
then
$\indmap{\cmap}$ is injective w.r.t. $D$-uniformity
for $\emptyset\neq D\subset\GF{p}$.
Based on the corresponding sufficient condition of entropy-preservation,
we construct three families of entropy-preserving maps and improve
previous results of \cite{SQ,TQ07,ZQ04,ZQ07it}.

The rest of this article is organized as follows.
In Sect. \ref{sect:main}, we present the formal definitions and
our main results.
In Sect. \ref{sect:pf-main},
we give the detailed proof for our main results.
In Sect. \ref{sect:conclusion}, we give a summary
and also comment on future work.

\section{Main results}\label{sect:main}
In this section we present our main results together
and leave the proof in Section \ref{sect:pf-main}.

Denote the set of rational integers by $\Int$. Assign $n,e\in\Int$
to be integers larger than one. Let $p$ be an odd prime, $\GF{p}$
the finite field of $p$ elements, $\uGF{p}=\GF{p}\setminus\set{0}$,
$\pring$ the residue ring modulo $p^e$, and $\gset$
the multiplicative group of $\pring$. For $r\in \pring$ and
a subset $A\subset \pring$, denote $r+A=\set{r+a:a\in A}$ and
$rA=\set{ra:a\in A}$. A \emph{sequence} $\seq{s}$ over $\pring$ is a
map $\seq{s}: \Int \rightarrow {\pring}$. Let $\seq{0}$ denote the
zero sequence.

Let $\prmpoly=x^\len+c_{\len-1}x^{\len-1}+\cdots+c_0$ be a monic
polynomial over $\pring$. Always suppose $\prmpoly \bmod p$ to be an
irreducible polynomial over $\GF{p}$. Let $\gseqg$ denote the set of
linear recurring sequences generated by $\prmpoly$ and
$\gseq=\set{\seq{s}\in\gseqg: \seq{s}\not\equiv\seq{0}\bmod p}$.
Without ambiguity we also write $x$ as the left-shift operator on sequences, 
i.e., $(\sum_{i=0}^k r_ix^i)\seq{s}$, where $r_i\in\pring$, is the sequence defined by
$(\sum_{i=0}^k r_ix^i)\seq{s}(t)=\sum_{i=0}^k r_i\seq{s}(t + i)$ for all $t\in\Int$.

It is known in the literature \cite{HD,kzm2} that there exist a polynomial $\delta$
over $\pring$ 
satisfying $\deg\delta<\len$
and $x^{p^\len-1} - 1 \equiv p\delta\bmod \prmpoly$.
\begin{definition}\label{def:primitivity}
The polynomial $\prmpoly$ is said to be \emph{primitive} if
$\prmpoly\bmod p$ is primitive over $\GF{p}$ and
$\delta\not\equiv0\bmod p$. The polynomial $\prmpoly$ is said to be
\emph{strongly primitive} if $\prmpoly\bmod p$ is primitive over
$\GF{p}$ and $(\delta\bmod p)\notin\GF{p}$.
A sequence $\seq{s}\in\gseq$ is said to be \emph{primitive} if
$\prmpoly$ is primitive.
\end{definition}

\begin{remark}\label{rk:strong-prim}
In definition \ref{def:primitivity}, we characterize (strong) primitivity
as in \cite{ward}, and primitivity is equivalent to
\begin{equation*}
\min\set{i>0: \prmpoly\mid x^i-1}=p^{e-1}\perb.
\end{equation*}
\end{remark}

Always let $\cmap:\pring\rightarrow \imset$ be a map defined on
$\pring$, and $\cmap$ is said to be \emph{constant on} $A \subset
\pring$ if $\cmap(a)=\cmap(b)$ for any $a,b\in A$.
For $A\subset \pring$,
denote $\cmap(A)=\set{\cmap(a):a\in A}$.
The map $\cmap$
naturally induces a map $\indmap{\cmap}$ on $\gseq$ defined by
$\indmap{\cmap}:\seq{s}\mapsto\cmap\comp\seq{s}$,
and $\indmap{\cmap}(\seq{s})$ is called a \emph{compressed sequence}.
\begin{definition}\label{def:uniformity}
Let $\seq{s}_1$ and $\seq{s}_2$ be two sequences over $\imset$, and
$D\subset \imset$. If for any $d\in D$ and $t\in\Int$, $\seq{s}_1(t)=d$ if and only
if $\seq{s}_2(t)=d$, then $\seq{s}_1$ and
$\seq{s}_2$ are called to be \emph{$D$-uniform}.
\end{definition}
\begin{remark}\label{rk:uniformity}
In Definition \ref{def:uniformity},
if we take $\imset=\GF{p}$ and $D=\set{s}$,
then $D$-uniformity is exactly
$s$-uniformity defined in \cite{JL14,ZxQ10,ZxQ13uni,ZQ08}.
\end{remark}
\begin{definition}\label{def:injectivity}
For $D\subset \imset$,
the induced map $\indmap{\cmap}$ is said to be \emph{injective on $\gseq$ w.r.t. $D$-uniformity}
if the following statement holds:
For $\seq{u},\seq{v}\in\gseq$,
$\indmap{\cmap}(\seq{u})$ and
$\indmap{\cmap}(\seq{v})$ are $D$-uniform if and only if
$\seq{u}=\seq{v}$.
\end{definition}
\begin{remark}\label{rk:injectivity}
(i) Injectivity w.r.t. $D$-uniformity means that the distribution of
elements of $D$ in the compressed sequence derived from $\cmap$
contains all the information of the original sequence. (ii)
For two sequences $\seq{s}_1$ and $\seq{s}_2$  over $\imset$,
$\seq{s}_1$ and $\seq{s}_2$ are $\imset$-uniform if and only if
$\seq{s}_1=\seq{s}_2$.
In 
\cite{qwf01},
the map $\cmap$ is called \emph{entropy-preserving} if
$\indmap{\cmap}$ is injective on $\gseq$.
Hence, entropy-preservation of $\cmap$ exactly
means that $\indmap{\cmap}$ is injective on $\gseq$ w.r.t. $\imset$-uniformity.
\end{remark}

Under the condition
$(\delta^2\bmod (p,\prmpoly))\notin\GF{p}$,
the following theorem gives an explicit criterion
of $D$-uniformity between primitive sequences.
\begin{theorem}\label{thm:result-uniform}
Let $\prmpoly$ be a primitive polynomial over $\pring$ satisfying
$\delta^2\not\equiv a \bmod (p,\prmpoly)$ for any $a\in\GF{p}$.
Let $\seq{u},\seq{v}\in\gseq$, $\ell=\max\set{0\leq i\leq
e:\exists j\in\pring, \seq{v}\equiv j\seq{u}\bmod p^i}$.
Suppose that $\gamma\in\pring$ satisfies $\seq{v}\equiv\gamma\seq{u}\bmod
p^\ell$.
Denote
\begin{equation*}
    \mathscr{C}=\left\{
    \begin{array}{l@{\text{ if }}l}
     \set{\pring}, & \ell=0,\\
     \set{\set{a\gamma^i:i\in\Int}+p^\ell\pring:a\in\pring}, & 1\leq \ell\leq e.
    \end{array}
    \right.
\end{equation*}
Then for  $D\subset \imset$, $\indmap{\cmap}(\seq{u})$ and
$\indmap{\cmap}(\seq{v})$ are $D$-uniform if and only if $\cmap$ is
constant on any set in $\set{C\in\mathscr{C}:\cmap(C)\cap D\neq\emptyset}$.
\end{theorem}

\begin{remark}
Given $\seq{u}(i)$ and $\seq{v}(i)$,
$t_0\leq i\leq t_0+\len-1$,
we can algorithmically compute $\ell$ and $\gamma$ of
Theorem \ref{thm:result-uniform} as below.
First, via the recurring relation Eq.(\ref{eqn:lfsr}) we get $\seq{u}(i)$,
$t_0+\len\leq i\leq t_0+2(\len-1)$. Denote the symmetric matrix
 \begin{equation*}
\Gamma = \begin{pmatrix}
     \seq{u}(t_0)&\seq{u}(t_0+1)&\cdots&\seq{u}(t_0+\len-1)\\
     \seq{u}(t_0+1)&\seq{u}(t_0+2)&\cdots&\seq{u}(t_0+\len)\\
     \vdots  & \vdots &\vdots  & \vdots \\
     \seq{u}(t_0+\len-1)&\seq{u}(t_0+\len)&\cdots&\seq{u}(t_0+2\len-2)\\
    \end{pmatrix}.
 \end{equation*}
Second, we compute a vector $(w_0,w_1,\dots,w_{\len-1})\in(\pring)^\len$
by solving
\begin{equation}\label{eqn:gamma-u-v}
(w_0,w_1,\dots, w_{\len-1})\Gamma
=(\seq{v}(t_0),\seq{v}(t_0+1),\dots,\seq{v}(t_0+\len-1)).
\end{equation}
Since $\seq{u}\bmod p$ is an $m$-sequence (by Lemma \ref{lemma:prim-seq-mod-p} later),
$\Gamma\bmod p$ is an invertible
matrix over $\GF{p}$ and hence Eq.(\ref{eqn:gamma-u-v}) is solvable.
Finally, noticing $\seq{v}=(w_0+ w_1x +\cdots +w_{\len-1}x^{\len-1})\seq{u}$,
we obtain $\ell=\min\set{k: p^k\mid w_i, 1\leq i\leq\len-1}$ and
$\gamma = w_0$.
\end{remark}

\begin{remark}\label{thm:special-lin-ind}
In Theorem \ref{thm:result-uniform},
$\ell=0$ means exactly that $\seq{u}\bmod p$ and
$\seq{v}\bmod p$ are linearly independent over $\GF{p}$.
Thus, for such sequences under the condition of Theorem \ref{thm:result-uniform},
 $D$-uniformity is explicitly described as follows:
%
For any nonempty $D\subset \imset$ with $D\cap\cmap(\pring)\neq\emptyset$,
$\indmap{\cmap}(\seq{u})$ and
$\indmap{\cmap}(\seq{v})$ are $D$-uniform if and only if $\cmap$ is
constant on $\pring$.
\end{remark}

The following two theorems give sufficient and equivalent conditions
on $\cmap$ such that $\indmap{\cmap}$ is injective w.r.t. $D$-uniformity.
\begin{theorem}\label{thm:uniform-branch-p-3}
Let $\prmpoly$ be a primitive polynomial over $\pring$ and
$\emptyset\neq D\subset \imset$.
Then $\indmap{\cmap}$ is injective on $\gseq$ w.r.t. $D$-uniformity
if the following three statements hold:
(i) For any $r$-th root of unity $1\neq\omega\in\pring$,
where $r$ is a prime divisor of $p-1$, there exists $a\in\gset$
($a\in{\pring}$ if $\prmpoly$ is strongly primitive) such that
$\cmap(a)\in D$ and $\cmap$ is not constant on $\set{a\omega^i:1\leq
i\leq r}$. (ii) There exists $a\in\gset$ such that $\cmap(a)\in D$
and  $\cmap$ is not constant on $a+p^{e-1}\pring$.
(iii) There exists $a\in{p\pring}$ such that $\cmap(a)\in D$ and
$\cmap$ is not constant on $a+p^{e-1}\pring$.
\end{theorem}

\begin{theorem}\label{thm:uniform-result-hyper-prim-p3}
Let $\prmpoly$ be a primitive polynomial over $\pring$ satisfying
$\delta^2\not\equiv a \bmod (p,\prmpoly)$ for any $a\in\GF{p}$.
Then $\indmap{\cmap}$ is injective on $\gseq$ w.r.t. $D$-uniformity
if and only if the following two statements hold: (i) For any
$r$-th root of unity $1\neq\omega\in\pring$, where $r$ is a prime
divisor of $p-1$, there exists $a\in\pring$ such that $\cmap(a)\in
D$ and $\cmap$ is not constant on $\set{a\omega^i:1\leq i\leq r}$.
(ii) There exists $a\in\gset$ such that $\cmap(a)\in D$ and
$\cmap$ is not constant on $a + p^{e-1}\pring$.
\end{theorem}

Due to Statement (ii) of  Remark \ref{rk:injectivity},
from Theorem \ref{thm:uniform-branch-p-3}
and Theorem \ref{thm:uniform-result-hyper-prim-p3}
we immediately derive the following two corollaries
to decide entropy-preservation of the compressing map $\cmap$.
\begin{corollary}\label{thm:branch-p-3}
Let $\prmpoly$ be a primitive polynomial over $\pring$. Then
$\indmap{\cmap}$ is  injective on $\gseq$ if the following
three statements hold: (i) For any $r$-th root of unity
$1\neq\omega\in\pring$, where $r$ is a prime divisor of $p-1$, there
exists $a\in\gset$ ($a\in{\pring}$ if $\prmpoly$ is strongly
primitive) such that $\cmap$ is not constant on
$\set{a\omega^i:1\leq i\leq r}$.
(ii) There exists $a\in\gset$ such
that $\cmap$ is not constant on $a+p^{e-1}\pring$.
(iii) There exists
$a\in{p\pring}$ such that $\cmap$ is not constant on
$a+p^{e-1}\pring$.
\end{corollary}

\begin{corollary}\label{thm:result-hyper-prim-p3}
Let $\prmpoly$ be a primitive polynomial over $\pring$ satisfying
$\delta^2\not\equiv a \bmod (p,\prmpoly)$ for any $a\in\GF{p}$.
Then $\indmap{\cmap}$ is injective on $\gseq$ if and only if the
following two statements hold: (i) For any $r$-th root of unity
$1\neq\omega\in\pring$, where $r$ is a prime divisor of $p-1$, there
exists $a\in\pring$ such that $\cmap$ is not constant on
$\set{a\omega^i:1\leq i\leq r}$. (ii) There exists $a\in\gset$
such that $\cmap$ is not constant on $a + p^{e-1}\pring$.
\end{corollary}

The following corollary shows that
if $\prmpoly$ is primitive and
$(\delta^2 \bmod (p,\prmpoly))\notin\GF{p}$,
then the majority of maps on $\pring$
are entropy-preserving. 
\begin{corollary}\label{cor:num-ep-map-p3}
Let $\imset$ be a finite set of cardinality $k$. If $\prmpoly$
is primitive and $\delta^2\not\equiv a \bmod (p,\prmpoly)$ for any
$a\in\GF{p}$, then the number of entropy-preserving maps from
$\pring$ to $\imset$ is greater than
\begin{equation*}
    k^{p^e}\left(1-k^{-(p-1)^2p^{e-2}}- k^{(1-p^e)/2}\log_2 p\right).
\end{equation*}
\end{corollary}

\begin{remark}\label{rk:hyper-prim-poly}
By Theorem \ref{fact:count-poly} in Subsect. \ref{subsect:no-prim-seq}, at least $1
- 2(p-1)/(p^\len-1)$ of $\len$-degree primitive polynomials 
satisfy the condition $(\delta^2\bmod (p,\prmpoly))\notin\GF{p}$
in Theorem \ref{thm:result-uniform},
Theorem \ref{thm:uniform-result-hyper-prim-p3},
Corollary \ref{thm:result-hyper-prim-p3}
and Corollary \ref{cor:num-ep-map-p3}.
The proportion $1 - 2(p-1)/(p^\len-1)$ approaches to $1$ as the
degree of polynomials increases.
\end{remark}

Now we consider some specific compressing maps.
We identify $a\in\pring$ with the vector
$(a_0,a_1,\dots,a_{e-1})\in\GF{p}^e$,
where 
$a=\sum_{j=0}^{e-1}a_j p^j$ and $a_i\in\set{0,1,\dots,p-1}$.
Then a map on $\pring$
is naturally written as an $e$-variable function over $\GF{p}$.

Based on Theorem \ref{thm:uniform-branch-p-3},
we give a family of compressing maps
such that the induced  maps on $\gseq$ are injective
w.r.t. $D$-uniformity.
\begin{theorem}\label{thm:uniform-perm-poly}
Let $\cmap(x)=f(x_{e-1})$, where $f(x_{e-1})$ is a permutation
polynomial over $\GF{p}$.
Then the following three statements are true:
\begin{enumerate}
  \item[(i)]
If $\prmpoly$ is a strongly primitive polynomial over $\pring$,
then for any $\emptyset\neq D\subset\GF{p}$,
$\indmap{\cmap}$ is injective on $\gseq$ w.r.t. $D$-uniformity.
  \item[(ii)]
If $\prmpoly$ is a primitive polynomial over $\pring$,
then for any $f(0)\in D\subset\GF{p}$,
$\indmap{\cmap}$ is injective on $\gseq$ w.r.t. $D$-uniformity.
\item[(iii)]
Let $\prmpoly$ be a primitive polynomial over $\pring$,
and $\seq{u},\seq{v}\in\gseq$ satisfy $\seq{u}\not\equiv-\seq{v}\bmod p^e$.
Then for any $\emptyset\neq D\subset\GF{p}$,
$\indmap{\cmap}(\seq{u})$ and $\indmap{\cmap}(\seq{v})$
are $D$-uniform if and only if $\seq{u}=\seq{v}$.
\end{enumerate}
\end{theorem}
\begin{remark}
The main result of \cite{ZQ05,ZQ07}
is a special case of Statement (ii)
 of Theorem \ref{thm:uniform-perm-poly}
with $D=\set{0}$ and $f(x_{e-1}) = x_{e-1}$.
\end{remark}

Furthermore, based on Corollary \ref{thm:branch-p-3}, we also
construct three families of entropy-preserving maps below.
\begin{theorem}\label{thm:new-map-str}
Let $\prmpoly$ be a strongly primitive polynomial over $\pring$.
A map $\cmap:\pring\rightarrow\GF{p}$ is written as
\begin{equation*}\label{eqn:new-map-str}
\cmap(x)=f_0(x_{e-1})f_1(x_0,x_1,\dots,x_{e-2})
+f_2(x_0,x_1,\dots,x_{e-2}),
\end{equation*}
where $f_0\in\GF{p}[x_{e-1}]$ and
$f_1,f_2\in\GF{p}[x_0,x_1,\dots,x_{e-2}]$. If $1\leq\deg f_0<p$,
$\left(x_{0}^{p-1}-1\right)\nmid f_1$ and $f_1(0,0,\dots,0)\neq0$,
then the induced map $\indmap{\cmap}$ is injective on $\gseq$.
\end{theorem}
\begin{remark}
A function like Eq.(\ref{eqn:map-zq1}) is a special case of Theorem
\ref{thm:new-map-str} with $f_1=1$, and hence Theorem
\ref{thm:new-map-str} improves the corresponding results in
\cite{TQ07,ZQ04,ZQ07it}.
\end{remark}
\begin{theorem}\label{thm:new-map-weak}
Let $\prmpoly$ be a primitive polynomial over $\pring$. A map $\cmap:\pring\rightarrow\GF{p}$
is of the form
\begin{equation*}\label{eqn:new-map-weak}
\cmap(x)=x_{e-1}^\ell
f_1(x_0,x_1,\dots,x_{e-2})+f_2(x_0,x_1,\dots,x_{e-2}),
\end{equation*}
where $2\leq\ell<p$ and $f_1,f_2\in\GF{p}[x_0,x_1\dots,x_{e-2}]$.
If $\left(x_{0}^{p-1}-1\right)\nmid f_1$ and $x_0\nmid f_1$,
then the induced map $\indmap{\cmap}$ is injective on $\gseq$.
\end{theorem}
\begin{remark}
A function like  Eq.(\ref{eqn:map-zq3}) is a special case of Theorem
\ref{thm:new-map-weak} with $f_1=1$. Thus, Theorem
\ref{thm:new-map-weak} improves the corresponding result in
\cite{ZQ04}.
\end{remark}
\begin{theorem}\label{thm:new-map-weak-sq}
Let $\prmpoly$ be a primitive polynomial over $\pring$.
A map $\cmap:\pring\rightarrow\GF{p}$ is of the form
\begin{equation*}\label{eqn:new-map-weak}
\cmap(x)=x_{e-1}\left(g_0(x_{k})+g_1(x_0,x_1,\dots,x_{k-1})\right)
+f_2(x_0,x_1,\dots,x_{e-2}),
\end{equation*}
where $g_0\in\GF{p}[x_k]$, $g_1\in\GF{p}[x_0,x_1\dots,x_{k-1}]$ and
$f_2\in\GF{p}[x_0,x_1\dots,x_{e-2}]$. Then the induced map
$\indmap{\cmap}$ is injective on $\gseq$ if $\gcd(p-1,\deg g_0+1)=1$
and
\begin{equation*}
\left\{
     \begin{array}{l@{\text{ if }}l}
1\leq\deg g_0<p, &
1\leq k\leq e-2,\\
\left(x_{0}^{p-1}-1\right)\nmid g_0 \text{ and } x_0\nmid g_0, &
k=0.
     \end{array}\right.
\end{equation*}
\end{theorem}
\begin{remark}
A function like Eq.(\ref{eqn:map-zq2}) is a special case of Theorem
\ref{thm:new-map-weak-sq} with $k=e-2$ and $\deg g_0\geq2$. Thus,
Theorem \ref{thm:new-map-weak-sq} improves the corresponding result
in \cite{SQ}.
\end{remark}

\section{Proof of main results}\label{sect:pf-main}
This section is organized as follows: In Subsect.
\ref{subsect:preliminary}, we prepare basic mathematical tools.
We study an equivalence relation of two primitive sequences
$\seq{u}$ and $\seq{v}$, and
discuss two cases respectively in Subsect. \ref{subsect:case-irrt}
and in Subsect. \ref{subsect:case-rt}.
In Subsect. \ref{subsect:pf-main-thm},
we prove Theorem \ref{thm:result-uniform},
Theorem \ref{thm:uniform-branch-p-3},
Theorem \ref{thm:uniform-result-hyper-prim-p3}
and Corollary \ref{cor:num-ep-map-p3}.
In Subsect. \ref{subsect:new-map},
we prove Theorem \ref{thm:uniform-perm-poly},
Theorem \ref{thm:new-map-str},
Theorem \ref{thm:new-map-weak} and
Theorem \ref{thm:new-map-weak-sq}.
In Subsect. \ref{subsect:no-prim-seq}
we give the number of primitive polynomials
satisfying the condition $(\delta^2\bmod (p,\prmpoly))\notin\GF{p}$.

In this section we need the following notations.
%
For $\seq{s}\in\gseqg$,
by $\delta\seq{s}(t)$ we always mean the value of the sequence
$\delta\seq{s}$ at clock cycle $t$ instead of the polynomial
$\delta\cdot\seq{s}(t)$.
For $\seq{u},\seq{v}\in\gseqg$, we denote $\seq{u}\lindep\seq{v}$
(resp. $\seq{u}\linind\seq{v}$) if
$\seq{u}\bmod p$ and $\seq{v}\bmod p$ are linearly dependent
(resp. independent) over $\GF{p}$.

\subsection{Preliminaries}\label{subsect:preliminary}
In this subsection we prepare some useful facts.
The first part is about binary relations, and
the second part is about primitive sequences over $\pring$.

\subsubsection{Equivalence relation}
Firstly, we introduce a binary relation on sequences
 to characterize uniformity.

A \emph{(binary) relation} $\rlt{R}$ on $\pring$ is a subset
of $\pring\times \pring$. The relation $\rlt{R}$ is \emph{reflexive}
if $\pair{a}{a}\in \rlt{R}$ for any $a \in \pring$; $\rlt{R}$ is
\emph{symmetric} if $\pair{a}{b}\in \rlt{R}$ implies $\pair{b}{a}\in
\rlt{R} $ for any $a,b\in \pring$; $\rlt{R}$ is \emph{transitive} if
$\pair{a}{b}\in\rlt{R}$ and $\pair{b}{c}\in \rlt{R}$ imply
$\pair{a}{c}\in \rlt{R}$ for any $a, b, c \in \pring$. The \emph{symmetric
closure} of $\rlt{R}$ is the smallest symmetric relation including
$\rlt{R}$. The \emph{transitive closure} of $\rlt{R}$ is the
smallest
 transitive relation including $\rlt{R}$.
A relation that is reflexive, symmetric and transitive is an
\emph{equivalence relation}. If $\rlt{R}$ is an equivalence relation
on $\pring$, then the \emph{equivalence class} of $a\in \pring$
w.r.t. $\rlt{R}$ is the set $\set{b:\pair{a}{b}\in \rlt{R}}$. The
\emph{equivalence closure} of $\rlt{R}$ is the smallest equivalence
relation including $\rlt{R}$.

Given $\seq{u},\seq{v}\in\gseqg$ and $1\leq i \leq e$, define a
relation on $\pring$:
\begin{equation*}
\rlab[i] = \set{\pair{a}{b}\in\pring\times\pring: \exists t\in\Int,
\seq{u}(t) \equiv a \bmod p^i, \seq{v}(t) \equiv b \bmod p^i}.
\end{equation*}
%
Denote the symmetric closure of $\rlab[i]$ by $\rlabs[i]$, i.e.,
\begin{equation*}
   \rlabs[i]  = \set{\pair{a}{b} :
    \pair{a}{b} \in \rlab[i]\text{ or }\pair{b}{a} \in \rlab[i]}.
\end{equation*}
Generally, $\rlab[i]$ is not necessarily an equivalence relation.
Denote the equivalence closure of $\rlab[i]$ by $\rlabe[i]$. The set
of equivalence classes $\set{\set{b:\pair{a}{b}\in
\rlabe}:a\in\pring}$ 
describes a partition of $\pring$.

The following facts follow from the definitions above.
\begin{fact}\label{fact:relation-recur}
If $\pair{a}{b}\in \rlab[i]$, then there exist $a'\in a +
p^i\pring$ and $b'\in b +p^i\pring$ satisfying
$\pair{a}{b}\in\rlab$.
\end{fact}
\begin{fact}\label{fact:relation-equiv}
For $a_0,b\in\pring$,  $\pair{a_0}{b}\in\rlabe[i]$ is equivalent
to $a_0=b$ or the fact that there exist
$\set{a_1,\dots,a_{k}}\subset\pring$ for some $k$ such that
$a_k=b$ and
$\pair{a_{j-1}}{a_j}\in\rlabs[i]$
for any $1\leq j\leq k$.
\end{fact}
\begin{fact}\label{fact:relation-subset-equiv-clss}
For $A\subset\pring$ and $1\leq i\leq e$, $A$ is a subset of an
equivalence class w.r.t. $\rlabe[i]$ if and only if $A\times
A\subset \rlabe[i]$.
\end{fact}

More about equivalence relations is available in \cite{gallier}.

The following lemma interprets $D$-uniformity via the binary relation $\rlabe$.
\begin{lemma}\label{lemma:map-eqrl}
Assume $\seq{u},\seq{v}\in\gseqg$ and let $\mathscr{C}$ be the set
of equivalence classes w.r.t. $\rlabe$. Then for any $\emptyset\neq D\subset
\imset$, $\indmap{\cmap}(\seq{u})$ and $\indmap{\cmap}(\seq{v})$ are
$D$-uniform if and only if $\cmap$ is constant on any set in
$\set{C\in\mathscr{C}:\cmap(C)\cap D\neq\emptyset}$. Particularly,
$\indmap{\cmap}(\seq{u})=\indmap{\cmap}(\seq{v})$ if and only if
$\cmap$ is constant on any $C\in\mathscr{C}$.
\end{lemma}
\pf Suppose $\cmap$ is constant on any $C\in\mathscr{C}$ with
$\cmap(C)\cap D\neq\emptyset$. Assume $\cmap(\seq{u}(t_0))\in D$ or
$\cmap(\seq{v}(t_0))\in D$ for some $t_0\in\Int$. Because
$\pair{\seq{u}(t_0)}{\seq{v}(t_0)}\in\rlab$ and
$\rlab\subset\rlabe$, $\seq{u}(t_0)$ and $\seq{v}(t_0)$ belong to
the same equivalence class w.r.t. $\rlabe$. Then
$\cmap(\seq{u}(t_0))=\cmap(\seq{v}(t_0))$. Thus,
$\indmap{\cmap}(\seq{u})$ and $\indmap{\cmap}(\seq{v})$ are
$D$-uniform.

Suppose $\indmap{\cmap}(\seq{u})$ and $\indmap{\cmap}(\seq{v})$ are
$D$-uniform.
Let  $C\in\mathscr{C}$ satisfy $\cmap(C)\cap D\neq\emptyset$. Choose any $a_0\in C$ with $\cmap(a_0)\in D$.
Assign any $b\in C$ with $b\neq a_0$.
By Fact
\ref{fact:relation-equiv}, there exist $a_1,\dots,a_{k}\in\pring$
for some $k$ such that $a_k=b$ and  $\pair{a_{j-1}}{a_j}\in\rlabs$
for any $1\leq j\leq k$. Then $D$-uniformity implies
$\cmap(a_0)=\cmap(a_1)=\cdots=\cmap(a_k)\in D$. Then we have $\cmap(a_0)=\cmap(b)$.
Thus, $\cmap$ is constant on $C$.

Therefore, $\indmap{\cmap}(\seq{u})$ and $\indmap{\cmap}(\seq{v})$
are $D$-uniform if and only if $\cmap$ is constant on any
$C\in\mathscr{C}$ with $\cmap(C)\cap D\neq\emptyset$.

Consider the particular case $D=\imset$. By Remark
\ref{rk:injectivity},
$\indmap{\cmap}(\seq{u})=\indmap{\cmap}(\seq{v})$ is equivalent to
$\imset$-uniformity of $\indmap{\cmap}(\seq{u})$ and
$\indmap{\cmap}(\seq{v})$, i.e.,  $\cmap$ is constant on any
$C\in\mathscr{C}$, since we always have $\cmap(C)\cap \imset\neq\emptyset$. \fp

By Lemma \ref{lemma:map-eqrl}, the key to deciding whether two
compressed sequences are $D$-uniform (or identical) is to characterize
the equivalence classes w.r.t. $\rlabe$.

The following two lemmas about $\rlabe$ are useful in the proof later.
\begin{lemma}\label{lemma:equiv-class-comb}
Let $A,B\subset\pring$ satisfying $A\times A\subset \rlabe$ and
$B\times B\subset\rlabe$. If there exist $a\in A$ and $b\in B$ with
$\pair{a}{b}\in\rlab$, then $(A\cup B)\times (A\cup B)\subset
\rlabe$.
\end{lemma}
\pf By Fact \ref{fact:relation-subset-equiv-clss}, we actually have
to show that $A\cup B$ is a subset of an equivalence class w.r.t.
$\rlabe$. Choose any $b'\in B$. Since $B\times B\subset \rlabe$, we
have $\pair{b}{b'}\in \rlabe$. Besides,
$\pair{a}{b}\in\rlab\subset\rlabe$.
Then we have $\pair{a}{b'}\in \rlabe$
by transitivity of $\rlabe$. Hence, $B\subset
\set{c\in\pring:\pair{a}{c}\in\rlabe}$. The condition $A\times
A\subset\rlabe$ implies $A\subset
\set{c\in\pring:\pair{a}{c}\in\rlabe}$. Thus, $A\cup B\subset
\set{c\in\pring:\pair{a}{c}\in\rlabe}$, that is $(A\cup B)\times
(A\cup B)\subset \rlabe$ by Fact
\ref{fact:relation-subset-equiv-clss}. \fp

\begin{lemma}\label{lemma:local-branch-irrt}
Let $1\leq \tau<e$ and $B\in\pring$ satisfying $B+p^{\tau}\pring =
B$. Assume that (i) for any $a_0,b\in B$ there exist $a_1,\dots,a_l$
for some $l$ such that $a_l = b$ and $\pair{a_{j-1}}{a_{j}}\in\rlab[\tau]$, $1\leq
j\leq l$;
and that (ii) for any $\tau<i\leq e$, $a_0\in B$ and $b\in a_0 +
p^{i-1}\pring$, there exist $a_1,\dots,a_{k}\in B$ for some
$k$ such that $a_k=b$ and $\pair{a_{j-1}}{a_{j}}\in\rlabs[i]$,
$1\leq j\leq k$. Then $B\times B\subset \rlabe$.
\end{lemma}
\pf
First, we use induction to prove the following statement:\\
\emph{Claim.} For any $\tau\leq i<e$ and $a\in B$, $a +p^{i}\pring$
is a subset of an equivalence class w.r.t. $\rlabe$, i.e., $(a
+p^{i}\pring) \times (a +p^{i}\pring) \subset \rlabe $.

By Assumption (ii) of Lemma \ref{lemma:local-branch-irrt}, for any $a=a_0\in B$ and $b\in
a_0+p^{e-1}\pring$, there exist $a_1,\dots,a_k\in B$ for
some $k$ satisfying $a_k=b$ and $\pair{a_{j-1}}{a_{j}}\in\rlabs$,
$1\leq j<k$. Due to the transitivity of $\rlabe$ and $\rlabs\subset
\rlabe$, we have $\pair{a}{b}\in\rlabe$. Hence, the claim holds for
$i=e-1$.

Suppose the claim is already proved for $i=i_0$, where  $\tau<i_0<e$. Below we show that the claim holds
for $i=i_0-1$.
Choose any $a_0\in B$ and $b\in a+p^{i_0-1}\pring$. Because of
Assumption (ii) of Lemma \ref{lemma:local-branch-irrt}, there exist $a_1,\dots,a_k\in B$ for some $k$
satisfying $b=a_k$ and $\pair{a_{j-1}}{a_{j}}\in\rlabs[i_0]$, $1\leq
j\leq k$. Denote $a_0^R=a_0$ and $a_k^L=a_k$. By Fact
\ref{fact:relation-recur}, for any $1\leq j\leq k$, there exist
$a_{j-1}^{L}$ and $a_j^R$ such that
$\pair{a_{j-1}^L}{a_{j}^R}\in\rlabs$, $1\leq j\leq k$, and
$a_j^L\equiv a_j\equiv a_j^R \bmod p^{i_0}$, $0\leq j\leq k$. Since
$B+p^\tau\pring=B$, we have $a_j^R,a_j^L\in B$, $0\leq j\leq k$. Because the claim is
proved for $i=i_0$, we have $\pair{a_j^R}{a_j^L}\in\rlabe$, $0\leq
j\leq k$. Due to transitivity of $\rlabe$ along each pair of
adjacent elements of the sequence
$a_0^R,a_0^L,a_1^R,a_1^L,\dots,a_k^R,a_k^L$, we get
$\pair{a_0^R}{a_k^L}\in\rlabe$, i.e., $\pair{a}{b}\in\rlabe$.
Therefore, for any $a\in\gset$, $a+p^{i_0-1}\pring$ is a subset of
an equivalence class w.r.t. $\rlabe$. Till now the claim is proved.

Now we use the claim to finish the proof.
Fix any $a_0,b\in B$. By Assumption (i) of Lemma \ref{lemma:local-branch-irrt}, there exist
$a_1,\dots,a_l\in B$ for some $l$ satisfying $a_l=b$ and
$\pair{a_{j-1}}{a_j}\in\rlab[\tau]$, $1\leq j\leq l$. By Fact
\ref{fact:relation-recur}, there exists $a_j^L,a_j^R\in
a_j+p^\tau\pring\subset B$, $0\leq j\leq l$, satisfying
$\pair{a_{j-1}^L}{a_j^R}\in\rlab$, $1\leq j\leq l$.
As shown in the claim above, $a_j +p^\tau\pring $ is a subset of an equivalence class w.r.t. $\rlabe$.
Iteratively using Lemma \ref{lemma:equiv-class-comb},
we conclude that
$\cup_{j=0}^l \left(a_j+p^\tau\pring\right)$ is a subset of an
equivalence class w.r.t. $\rlabe$ and hence
$\pair{a_0}{b}\in\rlabe$. Therefore, $B\times B\subset \rlabe$. \fp

\subsubsection{Sequences over rings}
Secondly, we also need the following properties of
primitive sequences.

\begin{lemma}\label{lemma:KKMN}
\cite[Theorem 26.1]{KKMN} Let $\seq{s}_1,\dots,\seq{s}_k$ are
linearly independent $m$-sequences over $\GF{p}$ generated by the
same primitive polynomial of degree $\len$. Then for
$v_1,\dots,v_k\in\GF{p}$,
\begin{equation*}
    \cset{\set{1\leq t\leq p^\len-1: \seq{s}_i(t)=v_i, 1\leq i\leq k }}=
    \left\{
    \begin{array}{ll}
        p^{\len-k}-1, & \text{ if } v_1=\cdots=v_k=0;\\
        p^{\len-k}, & \text{ otherwise}.
    \end{array}
    \right.
\end{equation*}
\end{lemma}

\begin{lemma}\label{lemma:prim-seq-mod-p}
If $\prmpoly$ is primitive and $\seq{s}\in\gseq$, then both
$\seq{s}\bmod p$ and $\delta\seq{s}\bmod p$ are $m$-sequences over
$\GF{p}$ generated by $\prmpoly\bmod p$.
\end{lemma}
\pf Because
\begin{equation*}
(\prmpoly\bmod p)(\seq{s}\bmod p)=\prmpoly\seq{s}\bmod
p=\seq{0}
\end{equation*}
and
\begin{equation*}
(\prmpoly\bmod p)(\delta\seq{s}\bmod p)=
\prmpoly\delta\seq{s} \bmod p = \delta (\prmpoly\seq{s})\bmod p =
\seq{0},
\end{equation*}
both $\seq{s}\bmod p$ and $\delta\seq{s}\bmod p$ are
generated by the primitive polynomial $\prmpoly \bmod p$. 
Besides, $\seq{s}\bmod p\neq \seq{0}$ and hence
$\seq{s}\bmod p$ is an $m$-sequence generated by $\prmpoly\bmod p$.
We also have $\delta\seq{s}\bmod p\neq\seq{0}$. Otherwise,
suppose $\delta\seq{s}\equiv\seq{0}\bmod p$ and then $\seq{s}\bmod
p$ is generated by $\delta\bmod p$ and its period is not larger than
$p^{\deg\delta}-1<p^{\deg\prmpoly}-1$, contradictory to the proved
fact that $\seq{s}\bmod p$ is an $m$-sequence generated by
$\prmpoly\bmod p$. Therefore, the supposition is absurd and
$\delta\seq{s}\bmod p$ is also an $m$-sequence generated by
$\prmpoly\bmod p$. \fp

\begin{lemma}\label{lemma:linear-indep}
Let $\seq{s},\seq{u},\seq{v}\in\gseq$. If $\prmpoly$ is primitive
but not strongly primitive, then $\seq{s}\lindep\delta\seq{s}$. If
$\prmpoly$ is strongly primitive, then
$\seq{s}\linind\delta\seq{s}$. If $\prmpoly$ is  primitive,
then $\seq{u}\lindep\seq{v}$ if and only if
$\delta\seq{u}\lindep\delta\seq{v}$.
\end{lemma}
\pf By Definition \ref{def:primitivity}, if $\prmpoly$ is primitive
but not strongly primitive, then $\delta\in\uGF{p}$ and hence
$\seq{s}\lindep\delta\seq{s}$.

Suppose $\prmpoly$ to be strongly primitive. If
$\seq{s}\lindep\delta\seq{s}$, say $\delta\seq{s} \equiv \lambda
\seq{s}\bmod p$ for some $\lambda\in\uGF{p}$, then
$(\delta-\lambda)\seq{s}\bmod p=\seq{0}$. By Lemma
\ref{lemma:prim-seq-mod-p},  $\seq{s}\bmod p$ is an $m$-sequence
generated by $\prmpoly$. Considering
$\deg(\delta-\lambda)<\deg\prmpoly$, we have
$\delta\equiv\lambda\bmod p$, contradictory to strong primitivity.
Thus, $\seq{s}\linind\delta\seq{s}$.

Suppose $\prmpoly$ to be primitive. If
$\seq{u}\lindep\seq{v}$, $\seq{u} \equiv \lambda \seq{v}\bmod p$ for
some $\lambda\in\uGF{p}$, then $\delta\seq{u} \equiv \lambda
\delta\seq{v}\bmod p$ and hence $\delta\seq{u}\lindep\delta\seq{v}$.
If $\delta\seq{u}\lindep\delta\seq{v}$, say $\delta\seq{u} \equiv
\lambda \delta\seq{v}\bmod p$ for some $\lambda\in\uGF{p}$, then
$\delta(\seq{u}-\lambda\seq{v})\bmod p=\seq{0}$. By Lemma
\ref{lemma:prim-seq-mod-p}, $(\seq{u}-\lambda\seq{v})\bmod p$ is
either $\seq{0}$ or an $m$-sequence generated by $\prmpoly$.
Considering $\deg\delta<\deg\prmpoly$, we have
$(\seq{u}-\lambda\seq{v})\bmod p=\seq{0}$, implying
$\seq{u}\lindep\seq{v}$. Therefore, $\seq{u}\lindep\seq{v}$ if and
only if
 $\delta\seq{u}\lindep\delta\seq{v}$.
\fp

\begin{lemma}\label{lemma:seq-p-adic}
Let $\seq{s}\in\gseqg$. Then for $1\leq i<e$ and $j\in\GF{p}$,
\begin{equation}\label{eqn:seq-shift-funda}
\seq{s}(t + jp^{i-1}\perb) \equiv \seq{s}(t) + jp^{i}
\delta\seq{s}(t) \mod p^{i+1}.
\end{equation}
\end{lemma}
\pf Since $x^{p^\len-1}-1 \equiv 1 + p\delta \bmod \prmpoly$, we
iteratively compute the $p$-th power of $(1+p\delta)^{jp^i}$,
$i\geq0$, and get
\begin{equation}\label{eqn:hop-p-adic}
    x^{jp^{i-1}\perb} \equiv 1 + j p^{i} \delta \mod (p^{i+1},\prmpoly),
\text{ for }i\geq1.
\end{equation}
Since $\seq{s}\in\gseqg$, using both hands of
Eq.(\ref{eqn:hop-p-adic}) as operators on $\seq{s}$ yields
Eq.(\ref{eqn:seq-shift-funda}). \fp

Now we consider which elements of $\pring$ occur in primitive
sequences.
\begin{lemma}\label{lemma:vl-exst}
Suppose $\prmpoly$ to be be primitive and
$\seq{s}\in\gseq$.
For any $a\in\pring$, if
there exist $t_0\in\Int$ satisfying $\seq{s}(t_0)\equiv a\bmod p$
and $\delta\seq{s}(t_0)\not\equiv0\bmod p$, then there exists
$t'\in t_0+\perb\Int$ satisfying $\seq{s}(t')= a$.
\end{lemma}
\pf For $1\leq i< e$, we iteratively take
\begin{equation*}
k_{i-1} = (a-\seq{s}(t_{i-1}))/
\left(p^{i}\delta\seq{s}(t_0)\right)\mod p
\end{equation*}
and $t_{i}=t_{i-1} + {p^{i-1}k_{i-1}}\perb$. Noticing $t_i\equiv t_{i-1}\equiv
t_0 \bmod p^\len-1$, by Lemma \ref{lemma:prim-seq-mod-p}, we have
$\delta\seq{s}(t_i)\equiv\delta\seq{s}(t_0)\not\equiv0\bmod p$. Then
for $i\geq2$, it follows from Eq.(\ref{eqn:seq-shift-funda}) that
\begin{equation*}
\seq{s}(t_{i})
 \equiv \seq{s}(t_{i-1})  + k_{i-1} p^{i}
 \delta\seq{s}(t_{i-1})
\equiv \seq{s}(t_{i-1})  + k_{i-1} p^{i}
 \delta\seq{s}(t_{0})
 \equiv  a \mod p^{i+1}.
\end{equation*}
Thus, $\seq{s}(t_i)\equiv a\bmod p^{i+1}$ and $k_{i}$ is
well-defined, $1\leq i< e$.
The proof is complete by taking $a'=a_{e-1}$.\fp

\begin{corollary}\label{cor:value-occur}
Let $\seq{s}\in \gseq$. If $\prmpoly$ is primitive, then
$\set{\seq{s}(t):t\in\Int}\supset \gset$. If $\prmpoly$ is strongly
primitive, then $\set{\seq{s}(t):t\in\Int}=\pring$.
\end{corollary}
\pf Suppose $\prmpoly$ to be strongly primitive and choose any $a\in
\pring$. By Lemma \ref{lemma:linear-indep}, $\seq{s}\bmod p$ and
$\delta\seq{s}\bmod p$ are linearly independent over $\GF{p}$. Then
by Lemma \ref{lemma:KKMN}, there exists $t_0$ satisfying
$\seq{s}(t_0)\equiv a\bmod p$ and $\delta\seq{s}(t_0) \not\equiv
0\bmod p$. Hence, by Lemma \ref{lemma:vl-exst}, there exists $t$
satisfying $\seq{s}(t)=a$. Thus, $\set{\seq{s}(t):t\in\Int}=\pring$.

Suppose $\prmpoly$ to be primitive but not strongly primitive and
choose any $a\in\gset$. In this case, $(\delta\bmod p)\in\uGF{p}$.
Since $\seq{s}\bmod p$ is an $m$-sequence by Lemma
\ref{lemma:prim-seq-mod-p}, there exists $t_0$ satisfying
$\seq{s}(t_0)\equiv a\bmod p$ and hence $\delta\seq{s}(t_0) \equiv
(\delta \bmod p)a \not\equiv 0\bmod p$. Then by Lemma
\ref{lemma:vl-exst}, there exists $t$ satisfying $\seq{s}(t)=a$.
Thus, $\set{\seq{s}(t):t\in\Int}\supset \gset$. \fp

As shown in the following example,
there exists a primitive sequence
$\seq{s}$ satisfying
$\set{\seq{s}(i):i\in\Int}\subsetneq\pring$.
\begin{example}\label{exmp:prim-p-3}
Take $e = p = 3$. The polynomial $\prmpoly = x^2  - x - 4$ over
$\Int/27\Int$ is primitive  but not strongly primitive because
$x^8 - 1 
\equiv -6\bmod (9,\prmpoly)$.
Let $\seq{s}\in\gseq$ satisfy
$\seq{s}(0)=2$ and
$\seq{s}(1)=13$. Then
\begin{equation*}
\set{\seq{s}(t):t\in\Int}=
\set{\pm1,\pm2,\pm3,\pm4,\pm5,\pm6,\pm7,\pm8,\pm10,\pm11,\pm12,
\pm13}.
\end{equation*}
\end{example}

\begin{remark}
For the case of strong primitivity, Corollary \ref{cor:value-occur}
can be proved by \cite[Lemma 16]{ZQ04}. Corollary
\ref{cor:value-occur} tells which elements of $\pring$ occur in
primitive sequences, and it is used in the proof
later. Here, we include it for completeness and readability.
\end{remark}

\subsection{Case I: On $\rlabe$ subject to $\seq{u}\linind\seq{v}$}
\label{subsect:case-irrt}
In this subsection let  $\seq{u},\seq{v}\in \gseq$ be two primitive
sequences satisfying $\seq{u}\linind\seq{v}$. We study the
equivalence class(es) w.r.t. $\rlabe$.

First we explicitly describe a condition here
because it will cause an irregular pit in our results.
\begin{condition}\label{cond:pit-p-lt-3}
It holds that
\begin{equation*}
    \left\{
    \begin{array}{l}
        ({\delta\bmod p})\notin\GF{p},\\
        \left(\delta^2\bmod (p,\prmpoly)\right) \in\uGF{p},\\
        \delta\seq{u}\lindep\seq{v}.
    \end{array}
    \right.
\end{equation*}
\end{condition}

\begin{remark}\label{rk:cond-pit-p3}
Under Condition \ref{cond:pit-p-lt-3}, $\delta\bmod p$ is a
quadratic non-residue in the extension of $\GF{p}$ by $\prmpoly\bmod
p$. Hence, we have $2\mid \deg\prmpoly$. Therefore, if the primitive
polynomial $\prmpoly$ is not strongly primitive or $2\nmid
\deg\prmpoly$, then Condition \ref{cond:pit-p-lt-3} does not hold.
\end{remark}

The following lemma relates Assumption (ii)
of Lemma \ref{lemma:local-branch-irrt}
to a system of equations.
\begin{lemma}\label{lemma:conn-irt-eqs}
Let $\seq{u},\seq{v}\in\gseq$ be primitive sequences satisfying
$\seq{u}\linind\seq{v}$. If there exists $t\in\Int$ satisfying
\begin{equation}\label{eqn:irrt-chk-p-3-2}
\left\{
\begin{array}{lcc}
   \seq{v}(t) & \equiv &  {a} \mod p,\\
   \delta\seq{v}(t) & \not\equiv & 0 \mod p, \\
   \delta\seq{u}(t) & \equiv & 0 \mod p,\\
   \seq{u}(t) & \not\equiv &  {0} \mod p,
\end{array}\right.
\end{equation}
or
\begin{equation}\label{eqn:irrt-chk-p-3-1}
\left\{
\begin{array}{lcc}
   \seq{u}(t)        & \equiv &  {a} \mod p ,\\
   \delta\seq{u}(t) &  \not\equiv & 0 \mod p, \\
   \delta\seq{v}(t) &  \equiv & 0\mod p,\\
   \seq{v}(t) & \not\equiv &  {0} \mod p,
\end{array}\right.
\end{equation}
then for any $2\leq i\leq e$, $a\in \gset$ and $b\in a + p^{i-1}\pring$,
there exist $c\in \gset$ such that  $\pair{a}{c}\in\rlabs[i]$ and
$\pair{c}{b}\in\rlabs[i]$.
\end{lemma}
\pf By Lemma \ref{lemma:vl-exst}, if there exists $t_0\in\Int$
satisfying Eq.(\ref{eqn:irrt-chk-p-3-2}), then there exists $t' \in
t_0 + \perb\Int$ satisfying $\seq{v}(t') = a$. Moreover, by Lemma
\ref{lemma:prim-seq-mod-p}, $\seq{u}\bmod p$, $\delta\seq{u}\bmod p$
and $\delta\seq{v}\bmod p$ are $m$-sequences generated by
$\prmpoly\bmod p$, implying
\begin{equation}\label{eqn:irrt-chk-p-3-2-impl}
     \left\{
     \begin{array}{lclc}
\seq{u}(t')      & \equiv & \seq{u}(t_0)      & \not\equiv0\mod p,\\
\delta\seq{u}(t')& \equiv & \delta\seq{u}(t_0)& \equiv 0 \mod p,\\
\delta\seq{v}(t')& \equiv & \delta\seq{v}(t_0)& \not\equiv 0 \mod p.
     \end{array}
     \right.
\end{equation}
On one hand, by Lemma \ref{lemma:seq-p-adic} and
Eq.(\ref{eqn:irrt-chk-p-3-2-impl}), for any $j\in\GF{p}$ we have
\begin{equation*}
\left\{
\begin{array}{l}
\seq{v}(t' + jp^{i-2}\perb) \equiv \seq{v}(t') +
jp^{i-1}\delta\seq{v}(t')
\equiv a + j p^{i-1}\delta\seq{v}(t')\mod p^i,\\
\seq{u}(t' + jp^{i-2}\perb) \equiv \seq{u}(t') +
jp^{i-1}\delta\seq{u}(t') \equiv  \seq{u}(t')\mod p^{i}.
\end{array}
\right.
\end{equation*}
On the other hand, $\pair{\seq{u}(t' + jp^{i-2}\perb)}{\seq{v}(t'
+ jp^{i-2}\perb)}\in\rlab$ for any $j\in\GF{p}$.
Thus, for any $b\in a + p^{i-1}\pring$, we have
$\pair{a}{\seq{u}(t')}\in\rlabs[i]$ and
$\pair{\seq{u}(t')}{b}\in\rlabs[i]$, where $\seq{u}(t')\in\gset$.

By Lemma \ref{lemma:vl-exst}, if there exists $t_0\in\Int$ satisfying
Eq.(\ref{eqn:irrt-chk-p-3-1}), then there exists $t' \in t_0 +
\perb\Int$ satisfying $\seq{u}(t') = a$. Moreover, by Lemma
\ref{lemma:prim-seq-mod-p}, $\seq{v}\bmod p$, $\delta\seq{v}\bmod p$
and $\delta\seq{u}\bmod p$ are $m$-sequences generated by
$\prmpoly\bmod p$, implying
\begin{equation}\label{eqn:irrt-chk-p-3-1-impl}
     \left\{
     \begin{array}{lclc}
\seq{v}(t')       &\equiv&\seq{v}(t_0)&\not\equiv0\mod p,\\
\delta\seq{v}(t') &\equiv &\delta\seq{v}(t_0)& \equiv 0 \mod p,\\
\delta\seq{u}(t') &\equiv &\delta\seq{u}(t_0) &\not\equiv 0 \mod p.
     \end{array}
     \right.
\end{equation}
On one hand,  by Lemma \ref{lemma:seq-p-adic}
and Eq.(\ref{eqn:irrt-chk-p-3-1-impl}), for
any $j\in\GF{p}$ we have
\begin{equation*}
\left\{
\begin{array}{l}
\seq{u}(t' + jp^{i-2}\perb) \equiv \seq{u}(t') +
jp^{i-1}\delta\seq{u}(t')
\equiv a + j p^{i-1}\delta\seq{u}(t') \mod p^i,\\
\seq{v}(t' + jp^{i-2}\perb) \equiv \seq{v}(t') +
jp^{i-1}\delta\seq{v}(t') \equiv \seq{v}(t') \mod p^i.
\end{array}
\right.
\end{equation*}
On the other hand, $\pair{\seq{u}(t' + jp^{i-2}\perb)}{\seq{v}(t'
+ jp^{i-2}\perb)}\in\rlab$ for any $j\in\GF{p}$.
Thus, for any $b\in a + p^{i-1}\pring$, we have
$\pair{a}{\seq{v}(t')}\in\rlabs[i]$ and
$\pair{\seq{v}(t')}{b}\in\rlabs[i]$,
where $\seq{v}(t')\in\gset$. \fp

In the following lemma we consider solvability of
Eq.(\ref{eqn:irrt-chk-p-3-2}) and Eq.(\ref{eqn:irrt-chk-p-3-1}).
\begin{lemma}\label{lemma:conn-irt-str}
Assume that $\prmpoly$ is strongly primitive and that Condition
\ref{cond:pit-p-lt-3} does not hold. If $\seq{u},\seq{v}\in\gseq$
and $\seq{u}\linind\seq{v}$, then for any $a\in\gset$, at least one
of Eq.(\ref{eqn:irrt-chk-p-3-2}) and Eq.(\ref{eqn:irrt-chk-p-3-1})
is satisfied.
\end{lemma}
\pf First, we prove the result
in any of the following five scenarios.
Fix any $a\in\gset$.
\begin{enumerate}
 \item[(i)] Suppose that $\seq{v}\bmod p$, $\delta\seq{v}\bmod p$,
$\delta\seq{u}\bmod p$ and $\seq{u}\bmod p$ are linearly independent
over $\GF{p}$. Then by Lemma \ref{lemma:KKMN}, both
Eq.(\ref{eqn:irrt-chk-p-3-2}) and Eq.(\ref{eqn:irrt-chk-p-3-1}) are
solvable. 
 \item[(ii)] Suppose that $\seq{v}\bmod p$, $\delta\seq{v}\bmod p$ and
$\delta\seq{u}\bmod p$ are linearly independent over $\GF{p}$, and
$\seq{u} \equiv d_0\seq{v} + d_1 \delta\seq{v} +
d_2\delta\seq{u}\bmod p$, where $d_0,d_1,d_2\in\GF{p}$.
Since $\seq{u}\linind\delta\seq{u}$ by Lemma \ref{lemma:linear-indep},
either $d_0$ or $d_1$ is nonzero here. We consider two
cases.

Case I:  $d_1=0$. Then $d_0\in\uGF{p}$. By Lemma \ref{lemma:KKMN},
there exists $t_0\in\Int$ satisfying
\begin{equation*}
\left\{
\begin{array}{lcc}
   \seq{v}(t_0) & \equiv &  {a} \mod p,\\
   \delta\seq{v}(t_0) & \not\equiv & 0 \mod p, \\
   \delta\seq{u}(t_0) & \equiv & 0 \mod p.
\end{array}\right.
\end{equation*}
Then 
$\seq{u}(t_0)\equiv d_0\seq{v}(t_0) +
d_1 \delta\seq{v}(t_0) + d_2\delta\seq{u}(t_0) \equiv d_0 a \bmod p$,
implying $\seq{u}(t_0)\not\equiv0\bmod p$. In this case,
Eq.(\ref{eqn:irrt-chk-p-3-2}) is solvable.

Case II: $d_1\neq0$.
Since $p\geq3$, we can choose
$b\in\GF{p}\setminus\set{0,-d_0a/d_1}$. By Lemma \ref{lemma:KKMN},
there exists $t_0\in\Int$ satisfying
\begin{equation*}
\left\{
\begin{array}{lcc}
   \seq{v}(t_0) & \equiv &  {a} \mod p,\\
   \delta\seq{v}(t_0) & \equiv & b \mod p, \\
   \delta\seq{u}(t_0) & \equiv & 0 \mod p.
\end{array}\right.
\end{equation*}
Then $\seq{u}(t_0)\equiv d_0\seq{v}(t_0) + d_1 \delta\seq{v}(t_0) +
d_2\delta\seq{u}(t_0) \equiv d_0 a + d_1 b \not\equiv0\bmod p$,
implying $\seq{u}(t_0)\not\equiv0\bmod p$. In this case,
Eq.(\ref{eqn:irrt-chk-p-3-2}) is solvable.
    \item[(iii)]
Assume that there exist $r'_0,r'_1\in\Int$ satisfying $\delta\seq{u}
\equiv r'_0\seq{v} + r'_1{\delta}\seq{v}\bmod p$, where $p\nmid
r'_1$. By Lemma \ref{lemma:linear-indep}, we have
$\delta\seq{u}\linind\delta\seq{v}$, implying $p\nmid r'_0$.
Since $\seq{v}\linind\delta\seq{v}$ by Lemma \ref{lemma:linear-indep},
now we consider two cases below.

Case I:  $\seq{u}\bmod p$, $\seq{v}\bmod p$ and
$\delta\seq{v}\bmod p$ are linearly independent over $\GF{p}$.
By Lemma \ref{lemma:KKMN}, there exists $t_0\in\Int$ satisfying
\begin{equation*}
\left\{
\begin{array}{lcl}
   \seq{v}(t_0) & \equiv &  {a} \mod p,\\
   \delta\seq{v}(t_0) & \equiv & -r'_0{a}/r'_1 \mod p, \\
   \seq{u}(t_0) & \not\equiv & 0 \mod p.
\end{array}\right.
\end{equation*}
Then $\delta\seq{u}(t_0)\equiv r'_0\seq{v}(t_0) +
r'_1{\delta}\seq{v}(t_0)\equiv 0\bmod p$. Notice that
$\delta\seq{v}(t_0)\equiv-r'_0{a}/r'_1\not\equiv0\bmod p$. Hence,
Eq.(\ref{eqn:irrt-chk-p-3-2}) is solvable.

Case II: $\seq{u}\equiv
d'_0\seq{v} + d'_1{\delta}\seq{v}\bmod p$ for some
$d'_0,d'_1\in\Int$. By Lemma \ref{lemma:KKMN}, there exists
$t_0\in\Int$ satisfying $\seq{v}(t_0)\equiv {a}\bmod p$ and
$\delta\seq{v}(t_0)\equiv -r'_0{a}/r'_1\not\equiv0\bmod p$, and we also have
$\delta\seq{u}(t_0)\equiv r'_0\seq{v}(t_0) +
r'_1{\delta}\seq{v}(t_0)\equiv 0\bmod p$.
Notice
\begin{equation*}
\begin{pmatrix}
\seq{u}\\
\delta\seq{u}
\end{pmatrix}
\equiv
    \begin{pmatrix}
d'_0&d'_1\\r'_0&r'_1
    \end{pmatrix}
\begin{pmatrix}
\seq{v}\\
\delta\seq{v}
\end{pmatrix}
\mod p.
\end{equation*}
By Lemma
\ref{lemma:linear-indep}, we have $\seq{u}\linind\delta\seq{u}$.
Hence, $d'_0r'_1\not\equiv d'_1r'_0\bmod p$. Then $\seq{u}(t_0) =
d'_0\seq{v}(t_0) + d'_1{\delta}\seq{v}(t_0) \equiv d'_0a -
d'_1r'_0{a}/r'_1\not\equiv0 \bmod p$. Thus,
Eq.(\ref{eqn:irrt-chk-p-3-2}) is solvable.
    \item[(iv)]
Suppose that $\seq{u}\bmod p$, $\delta\seq{u}\bmod p$ and
$\delta\seq{v}\bmod p$ are linearly independent over $\GF{p}$, and
$\seq{v} \equiv d_0\seq{u} + d_1 \delta\seq{u} +
d_2\delta\seq{v}\bmod p$. Since $\seq{v}\linind\delta\seq{v}$ by Lemma \ref{lemma:linear-indep},
either $d_0$ or $d_1$ is nonzero here. We consider two
cases.

Case I:  $d_1=0$. Then $d_0\in\uGF{p}$. By Lemma \ref{lemma:KKMN},
there exists $t_0\in\Int$ satisfying
\begin{equation*}
\left\{
\begin{array}{lcc}
   \seq{u}(t_0) & \equiv &  {a} \mod p,\\
   \delta\seq{u}(t_0) & \not\equiv & 0 \mod p, \\
   \delta\seq{v}(t_0) & \equiv & 0 \mod p.
\end{array}\right.
\end{equation*}
Then $\seq{v}(t_0)\equiv d_0\seq{u}(t_0) +
d_1 \delta\seq{u}(t_0) + d_2\delta\seq{v}(t_0) \equiv d_0 a \bmod p$,
implying $\seq{v}(t_0)\not\equiv0\bmod p$. In this case,
Eq.(\ref{eqn:irrt-chk-p-3-1}) is solvable.

Case II: $d_1\neq0$.
Since $p\geq3$, we can choose
$b\in\GF{p}\setminus\set{0,-d_0a/d_1}$. By Lemma \ref{lemma:KKMN},
there exists $t_0\in\Int$ satisfying
\begin{equation*}
\left\{
\begin{array}{lcc}
   \seq{u}(t_0) & \equiv &  {a} \mod p,\\
   \delta\seq{u}(t_0) & \equiv & b \mod p, \\
   \delta\seq{v}(t_0) & \equiv & 0 \mod p.
\end{array}\right.
\end{equation*}
Then $\seq{v}(t_0)\equiv d_0\seq{u}(t_0) + d_1 \delta\seq{u}(t_0) +
d_2\delta\seq{v}(t_0) \equiv d_0 a + d_1 b \not\equiv0\bmod p$,
implying $\seq{v}(t_0)\not\equiv0\bmod p$. In this case,
Eq.(\ref{eqn:irrt-chk-p-3-1}) is solvable.
     \item[(v)]
Assume that there exist $r_0,r_1\in\Int$ satisfying $\delta\seq{v}
\equiv r_0\seq{u} + r_1{\delta}\seq{u}\bmod p$, where $p\nmid r_1$.
By Lemma \ref{lemma:linear-indep}, we have
$\delta\seq{u}\linind\delta\seq{v}$, implying $p\nmid r_0$.
Since $\seq{u}\linind\delta\seq{u}$ by Lemma \ref{lemma:linear-indep},
now we consider two cases.

Case I:  $\seq{v}\bmod p$, $\seq{u}\bmod p$ and
$\delta\seq{u}\bmod p$ are linearly independent over $\GF{p}$. Then
by Lemma \ref{lemma:KKMN}, there exists $t_0\in\Int$ satisfying
\begin{equation*}
\left\{
\begin{array}{lcl}
   \seq{u}(t_0) & \equiv &  {a} \mod p,\\
   \delta\seq{u}(t_0) & \equiv & -r_0{a}/r_1 \mod p, \\
   \seq{v}(t_0) & \not\equiv & 0 \mod p.
\end{array}\right.
\end{equation*}
Then $\delta\seq{v}(t_0)\equiv r_0\seq{u}(t_0) +
r_1{\delta}\seq{u}(t_0)\equiv 0\bmod p$. Notice that
$\delta\seq{u}(t_0)\equiv-r_0{a}/r_1\not\equiv0\bmod p$. Hence,
Eq.(\ref{eqn:irrt-chk-p-3-1}) is solvable.

Case II: $\seq{v}\equiv
d_0\seq{u} + d_1{\delta}\seq{u}\bmod p$ for some $d_0,d_1\in\Int$.
By Lemma \ref{lemma:KKMN}, there exists $t_0\in\Int$ satisfying
$\seq{u}(t_0)\equiv {a}\bmod p$ and $\delta\seq{u}(t_0)\equiv
-r_0{a}/r_1\not\equiv0\bmod p$, and we also have $\delta\seq{v}(t_0)\equiv
r_0\seq{u}(t_0) + r_1{\delta}\seq{u}(t_0)\equiv 0\bmod p$.
Notice
\begin{equation*}
\begin{pmatrix}
\seq{v}\\
\delta\seq{v}
\end{pmatrix}
\equiv
    \begin{pmatrix}
d_0&d_1\\r_0&r_1
    \end{pmatrix}
\begin{pmatrix}
\seq{u}\\
\delta\seq{u}
\end{pmatrix}
\mod p.
\end{equation*}
By Lemma
\ref{lemma:linear-indep}, we have $\seq{v}\linind\delta\seq{v}$.
Hence, $d_0r_1\not\equiv d_1r_0\bmod p$. Then $\seq{v}(t_0) =
d_0\seq{u}(t_0) + d_1{\delta}\seq{u}(t_0) \equiv d_0a -
d_1r_0{a}/r_1\not\equiv0 \bmod p$. Thus,
Eq.(\ref{eqn:irrt-chk-p-3-1}) is solvable.
\end{enumerate}

Recalling $\seq{v}\linind\delta\seq{v}$ and
$\seq{u}\linind\delta\seq{u}$, we conclude that if none of the five
scenarios (i)-(v) holds, then $r_1\equiv r'_1\equiv0\bmod p$, i.e.,
$\delta\seq{u}\lindep\seq{v}$ and $\delta\seq{v}\lindep\seq{u}$.
Under this condition, notice that
$\seq{v}\lindep\delta\seq{u}\lindep \delta(\delta\seq{v}) \lindep
\delta^2\seq{v}$. By Lemma \ref{lemma:prim-seq-mod-p}, $\seq{v}\bmod
p$  is an $m$-sequence generated by $\prmpoly\bmod p$. Then
$\left(\delta^2\bmod (p,\prmpoly)\right) \in\uGF{p}$. Thus,
Condition \ref{cond:pit-p-lt-3} holds if none of scenarios (i)-(v)
holds.

Therefore, if $\prmpoly$ is strongly primitive and Condition
\ref{cond:pit-p-lt-3} does not hold, then at least one of
the five scenarios (i)-(v) is true, and hence either
Eq.(\ref{eqn:irrt-chk-p-3-2}) or Eq.(\ref{eqn:irrt-chk-p-3-1}) is
solvable for any  $a\in\gset$.\fp

Combining Lemma \ref{lemma:local-branch-irrt},
Lemma \ref{lemma:conn-irt-eqs} and Lemma \ref{lemma:conn-irt-str},
we can determine the equivalence class w.r.t.
$\rlabe$ subject to some weak restrictions.
\begin{lemma}\label{lemma:equiv-class-irrt}
Let $\seq{u},\seq{v}\in\gseq$ and $\seq{u}\linind\seq{v}$. If
$\prmpoly$ is strongly primitive and Condition \ref{cond:pit-p-lt-3}
does not hold, then $\pring$ is the equivalence class w.r.t.
$\rlabe$.
\end{lemma}
\pf
First, we show $\gset\times \gset\subset\rlabe$. Clearly,
$\gset+p\pring=\gset$. Since $\seq{u}\linind\seq{v}$, by Lemma
\ref{lemma:KKMN}, for any $a,b\in \gset$, there exists $t\in\Int$
satisfying $\seq{u}(t)\equiv a\bmod p$ and $\seq{v}(t)\equiv b\bmod
p$. Then $\pair{a}{b}\in\rlab[1]$. Thus, Assumption (i) of Lemma
\ref{lemma:local-branch-irrt} holds  for $B = \gset$ and $\tau=1$.
Furthermore, by Lemma \ref{lemma:conn-irt-str} and Lemma
\ref{lemma:conn-irt-eqs}, Assumption (ii) of Lemma
\ref{lemma:local-branch-irrt} holds  for $B = \gset$ and $\tau=1$.
Therefore, it follows from Lemma \ref{lemma:local-branch-irrt} that
$\gset\times\gset\subset\rlabe$, i.e., $\gset$ is a subset of an
equivalence class w.r.t. $\rlabe$.

Now assign any $a\in p\pring$. We consider
\begin{equation}\label{eqn:conn-irt-0-1}
\left\{
\begin{array}{lcc}
   \seq{u}(t) & \equiv &  0 \mod p,\\
   \delta\seq{u}(t) & \not\equiv  & 0 \mod p, \\
   \seq{v}(t) & \not\equiv & 0 \mod p.
\end{array}\right.
\end{equation}
If $\seq{u}\bmod p$, $\delta\seq{u}\bmod p$ and $\seq{v}\bmod p$ are
linearly independent over $\GF{p}$, then Eq.(\ref{eqn:conn-irt-0-1}) is solvable
by Lemma \ref{lemma:KKMN}. Otherwise, suppose that
$\seq{u}\bmod p$, $\delta\seq{u}\bmod p$ and $\seq{v}\bmod p$ are
linearly dependent over $\GF{p}$. Since
$\seq{u}\linind\delta\seq{u}$ and $\seq{u}\linind\seq{v}$, we have
$\seq{v}\equiv d_0\seq{u}+d_1\delta\seq{u}\bmod p$ for some
$d_0,d_1\in\Int$, where $p\nmid d_1$. By Lemma \ref{lemma:KKMN},
there exists $t_0\in \Int$ satisfying   $\seq{u}(t_0) \equiv   0 \bmod
p$ and $  \delta\seq{u}(t_0) \not\equiv  0 \bmod p$. Then
$\seq{v}(t_0)\equiv d_1\delta\seq{u}(t_0)\not\equiv0\bmod p$. Thus,
Eq.(\ref{eqn:conn-irt-0-1}) is solvable in either case.
Denote $t_0$ to be a solution of Eq.(\ref{eqn:conn-irt-0-1}). By Lemma
\ref{lemma:vl-exst}, there exist $t'\in t_0+ \perb\Int$ such that
$\seq{u}(t')=a$. By Lemma \ref{lemma:prim-seq-mod-p},
 $\seq{v}(t')\equiv\seq{v}(t_0)\not\equiv0\bmod p$, i.e., $\seq{v}(t')\in\gset$.
Thus, $\pair{a}{\seq{v}(t')}\in\rlab$.

We have shown that for any
$a\in p\pring$, $\pair{a}{b}\in\rlab$ for some $b\in\gset$. Since
$\pring = (p\pring) \cup \gset$, by Lemma
\ref{lemma:equiv-class-comb}, we have $\pring\times
\pring\subset\rlabe$, i.e., $\pring$ is the equivalence class w.r.t.
$\rlabe$. \fp

Below an example under Condition \ref{cond:pit-p-lt-3} shows that
$\pring$ is not necessarily an equivalence class
even if $\prmpoly$ is strongly primitive.
\begin{example}
Take $p = 3$, $e = 2$, and $\prmpoly= x^2 + x - 1$ over
$\pring[9]$. Then $(x^8-1)/3 \equiv 1 -x \bmod (3,\prmpoly)$ and
$(x^8-1)^2/9 \equiv  2\bmod (3,\prmpoly)$. Let
$\seq{u},\seq{v}\in\gseq$ with $\seq{u}(0)=2$, $\seq{u}(1)=0$,
$\seq{v}(0)=1$ and $\seq{v}(1)=2$. Then the equivalence classes
w.r.t. $\rlabe$ are as follows:
\begin{equation*}
{\set{\pm4},\set{0,\pm1,\pm2, \pm3}}.
\end{equation*}
\end{example}

When Condition \ref{cond:pit-p-lt-3} holds
or $\prmpoly$ is not strongly primitive,
in the following two lemmas we find part information of
$\rlabe$.
\begin{lemma}\label{lemma:conn-irt-cnd1}
Assume that $\prmpoly$ is primitive and that Condition
\ref{cond:pit-p-lt-3} holds. Let $\seq{u},\seq{v}\in\gseq$ and
$\seq{u}\linind\seq{v}$. Then for any  $a\in p\pring$, $(a +
p^{e-1}\pring)\times (a + p^{e-1}\pring)\subset\rlabe$.
\end{lemma}
\pf Assign any $a\in p\pring$. Because $(\delta^2\bmod
(p,\prmpoly))\in\uGF{p}$ and $\delta\seq{u}\lindep\seq{v}$, we have
$\delta\seq{v}\lindep\delta(\delta\seq{u})\lindep\delta^2\seq{u}\lindep\seq{u}$.
Besides, $a\equiv0\bmod p$. Thus, Eq.(\ref{eqn:irrt-chk-p-3-2}) is
equivalent to
\begin{equation}\label{eqn:irrt-chk-p-3-cnd}
\left\{
\begin{array}{lcc}
   \seq{v}(t) & \equiv &  0 \mod p,\\
   \delta\seq{v}(t) & \not\equiv & 0 \mod p.
\end{array}\right.
\end{equation}
Notice that $\prmpoly$ is strongly primitive under Condition
\ref{cond:pit-p-lt-3}. By Lemma \ref{lemma:linear-indep},
$\seq{v}\linind\delta\seq{v}$. Then by Lemma \ref{lemma:KKMN},
Eq.(\ref{eqn:irrt-chk-p-3-cnd}) is solvable. Thus, by Lemma
\ref{lemma:conn-irt-eqs}, for any $b\in a + p^{e-1}\pring$, there
exists $c\in\pring$ such that $\pair{a}{c}\in\rlabs$ and
$\pair{c}{b}\in\rlabs$, implying $\pair{a}{b}\in\rlabe$. Therefore,
$(a+p^{e-1}\pring)\times(a+p^{e-1}\pring)\subset\rlabe$. \fp

\begin{lemma}\label{lemma:conn-irt-prm}
Assume that $\prmpoly$ is primitive but not strongly primitive. Let
$\seq{u},\seq{v}\in\gseq$ and $\seq{u}\linind\seq{v}$. Then for any
$a\in \gset$, $(a + p^{e-1}\pring)\times(a +
p^{e-1}\pring)\subset\rlabe$.
\end{lemma}
\pf Assign any $a\in\gset$, i.e., $a\not\equiv0\bmod p$. By Lemma
\ref{lemma:KKMN}, since $\seq{u}\linind\seq{v}$, there exists
$t_0\in\Int$ satisfying
\begin{equation}\label{eqn:irrt-chk-p-3-weak}
\left\{
\begin{array}{lcl}
  \seq{u}(t_0) & \equiv &  {a} \not\equiv 0 \mod p,\\
  \seq{v}(t_0) & \equiv & 0 \mod p.
\end{array}\right.
\end{equation}
By Lemma \ref{lemma:linear-indep},
$\delta\seq{u}\lindep\seq{u}$ and $\delta\seq{v}\lindep\seq{v}$. By Lemma \ref{lemma:prim-seq-mod-p},
$\delta\seq{u}\bmod p\neq \seq{0}$ and $\delta\seq{v}\bmod p\neq
\seq{0}$. Hence, $\delta\seq{u}(t_0)\not\equiv0\bmod p$ and
$\delta\seq{v}(t_0)\equiv0\bmod p$. On one hand,  it follows from
Lemma \ref{lemma:seq-p-adic} that for any $j\in\GF{p}$,
\begin{equation*}
\left\{
\begin{array}{l}
\seq{u}(t_0 + jp^{e-2}\perb) \equiv \seq{u}(t_0) +
jp^{e-1}\delta\seq{u}(t_0)
\equiv a + j p^{e-1}\delta\seq{u}(t_0) \mod p^e,\\
\seq{v}(t_0 + jp^{e-2}\perb) \equiv \seq{v}(t_0) +
jp^{e-1}\delta\seq{v}(t_0) \equiv \seq{v}(t_0) \mod p^e.
\end{array}
\right.
\end{equation*}
On the other hand, $\pair{\seq{u}(t_0 + jp^{e-2}\perb)}{\seq{v}(t_0
+ jp^{e-2}\perb)}\in\rlab$ for any $j\in\GF{p}$. Thus, for any $b\in
a + p^{e-1}\pring$, we have $\pair{a}{\seq{v}(t_0)}\in\rlab$ and
$\pair{\seq{v}(t_0)}{b}\in\rlabs$, implying $\pair{a}{b}\in\rlabe$.
Therefore, $(a + p^{e-1}\pring)\times(a +
p^{e-1}\pring)\subset\rlabe$. \fp

\subsection{Case II: On $\rlabe$ subject to $\seq{u}\lindep\seq{v}$}
\label{subsect:case-rt}
In this subsection let  $\seq{u},\seq{v}\in \gseq$ 
satisfying $\seq{u}\lindep\seq{v}$, $\ell =
\max\set{0\leq i\leq e:\exists j\in \Int, j\seq{u}\equiv
\seq{v}\bmod p^i}$ and
let $\gamma\in\pring$ satisfy $\gamma\seq{u}\equiv
\seq{v}\bmod p^\ell$.
We study the equivalence classes w.r.t. $\rlabe$.

In our proof we need the following relation
determined by $\seq{u}$ and $\seq{v}$.
For $1\leq i\leq e$, we define 
\begin{align*}
    \brlab[i]=&\{(a,b)\in\pring\times\pring:\exists t\in \Int,
    \delta\seq{u}(t)\not\equiv 0\bmod p,\\
    &\quad \seq{u}(t)\equiv a\bmod p^i,
    \seq{v}(t)\equiv b\bmod p^i
    \}.
\end{align*}
In the rest of this subsection, we denote
\begin{equation*}
    \ggset =\left\{
    \begin{array}{l@{\text{ if }}l}
      \pring, & \prmpoly \text{ is strongly primitive}, \\
      \gset, & \prmpoly \text{ is primitive but not strongly primitive}.
    \end{array}\right.
\end{equation*}

We relate the assumptions of Lemma \ref{lemma:local-branch-irrt}
to the following statement.
\begin{statement}\label{stat:strong-trans-rt}
For any $a\in\ggset$ and $b\in a + p^{l-1}\pring$, there exist $k>0$
and $\set{a_0,a_1,\dots,a_k}\subset\set{a\gamma^i:i\in\Int}+p^\ell\pring$
such that
\begin{equation*}
    \left\{
     \begin{array}{l}
       a_0 \equiv a \mod p^l, \\
       a_k \equiv b \mod p^l, \\
       \gamma^k\equiv1 \mod p^\ell,\\
       \pair{a_{j-1}}{a_j}\in\brlab[l], 1\leq j\leq k.
     \end{array}\right.
\end{equation*}
\end{statement}

Via the following three lemmas,
we use induction to prove Statement \ref{stat:strong-trans-rt}.
\begin{lemma}\label{lemma:w-u-v-lin-ind}
Let $\seq{u}$, $\seq{v}$, $\ell$ and $\gamma$ be defined as above
and $1\leq \ell<e$.
Let $\seq{w}=(\seq{v}-\gamma\seq{u})/p^\ell$. Then
$\seq{w}\linind\seq{u}$, $\seq{w}\linind\seq{v}$ and
$\seq{w}\bmod p$ is an $m$-sequence generated by $\prmpoly\bmod p$.
\end{lemma}
\pf Assume $\seq{w}\lindep\seq{u}$, i.e.,
$\seq{w}\equiv \lambda\seq{u}\bmod p$ for some $\lambda\in\Int$.
Then $\seq{v}\equiv \gamma\seq{u}+p^\ell\seq{w}
\equiv \gamma\seq{u}+\lambda p^\ell\seq{u}
\equiv (\gamma+\lambda p^\ell)\seq{u}\bmod p^{\ell+1}$,
contradictory to the definition of $\ell$. Therefore, our assumption is absurd and
$\seq{w}\linind\seq{u}$.

The proof of $\seq{w}\linind\seq{v}$ is similar.

As $\seq{w}\bmod p$ is linearly independent with the $m$-sequence
$\seq{u}\bmod p$,
it is necessary that $\seq{w}\not\equiv0\bmod p$.
Since
\begin{equation*}
p^\ell\prmpoly\seq{w}\equiv\prmpoly(p^\ell\seq{w})
\equiv \prmpoly(\seq{v}-\gamma\seq{u})\equiv
(\prmpoly\seq{v}-\gamma\prmpoly\seq{u})\equiv\seq{0}\bmod p^e,
\end{equation*}
we have
\begin{equation*}
(\prmpoly\bmod p)(\seq{w}\bmod p)\equiv
\prmpoly\seq{w}\equiv \seq{0}\bmod p.
\end{equation*}
Thus, $\seq{w}\bmod p$ is an $m$-sequence generated by $\prmpoly\bmod p$.
\fp

\begin{lemma}\label{lemma:rt-conn-ell}
Let $\seq{u}$, $\seq{v}$, $\ell$ and $\gamma$ be defined as above,
and $1\leq \ell < e$.
If $\prmpoly$ is primitive, then Statement \ref{stat:strong-trans-rt}
holds for $l=\ell + 1$.
\end{lemma}
\pf Let $\seq{w}=(\seq{v}-\gamma\seq{u})/p^\ell$.

Choose any $a\in \ggset$. Denote
\begin{equation*}
D_{{a}} = \set{\seq{w}(t)\bmod p:\exists t\in \Int,
       \seq{u}(t)\equiv{a}\bmod p,
       \delta\seq{u}(t)\not\equiv 0}.
\end{equation*}
By Lemma \ref{lemma:vl-exst}, for any $t\in\Int$ satisfying
$\seq{u}(t)\equiv{a}\bmod p$ and $\delta\seq{u}(t)\not\equiv 0\bmod
p$, there exists $t'\in t + \perb\Int$ with $\seq{u}(t')=a$. By
Lemma \ref{lemma:w-u-v-lin-ind},
$\seq{w}\bmod p$ is an $m$-sequence
generated by $\prmpoly\bmod p$. Then $\seq{v}(t') \equiv
\gamma\seq{u}(t') + p^\ell \seq{w}(t') \equiv \gamma a +
p^\ell(\seq{w}(t)\bmod p)\bmod p^{\ell+1}$.
By Lemma \ref{lemma:prim-seq-mod-p}, $\delta\seq{u}(t')\equiv\delta\seq{u}(t)\not\equiv0\bmod p$.
Since $\pair{\seq{u}(t')}{\seq{v}(t')}\in\rlab$,
we have $\pair{a}{\gamma a +p^\ell c} \in\brlab[\ell+1]$ for any $c\in D_a$.

Now we consider three cases.
\begin{enumerate}
 \item[(i)]
$\prmpoly$ is primitive but not strongly primitive.
Recall $(\delta\bmod p)\in\uGF{p}$ and $a\in\ggset=\gset$.
By Lemma \ref{lemma:linear-indep}, $\delta\seq{u}\lindep\seq{u}$
and hence
$D_{{a}}=\set{\seq{w}(t)\bmod p:
       \seq{u}(t)\equiv{a}\bmod p}$.
By Lemma \ref{lemma:w-u-v-lin-ind},
$\seq{w}\linind\seq{u}$.
Thus, we use Lemma \ref{lemma:KKMN} to get
$D_{{a}}=\set{\seq{w}(t)\bmod p:
       \seq{u}(t)\equiv{a}\bmod p}
       =\GF{p}$.
Thus, $\pair{a}{\gamma a+jp^\ell}\in\brlab[\ell+1]$ for $j\in\Int$.

In Cases (ii) and (iii) below,
$\prmpoly$ is strongly primitive.
\item[(ii)]  By Lemma \ref{lemma:KKMN}, if $\seq{u}\bmod p$,
${\delta}\seq{u}\bmod p$ and $\seq{w}\bmod p$ are linearly
independent over $\GF{p}$, then $D_{{a}}=\GF{p}$. Thus,
$\pair{a}{\gamma a+jp^\ell}\in\brlab[\ell+1]$ for $j\in\Int$.
\item[(iii)]
Otherwise, $\seq{u}\bmod p$, ${\delta}\seq{u}\bmod p$ and
$\seq{w}\bmod p$ are linearly dependent over $\GF{p}$.
By Lemma \ref{lemma:linear-indep},
$\seq{u}\linind\delta\seq{u}$. Suppose $\seq{w}\equiv
r_0\seq{u}+r_1\delta\seq{u}\bmod p$. By Lemma
\ref{lemma:w-u-v-lin-ind}, we have $p\nmid r_1$.
By Lemma
\ref{lemma:prim-seq-mod-p}, both $\seq{u}\bmod p$
and $\delta\seq{u}\bmod p$ are $m$-sequences over $\GF{p}$.
Then by Lemma \ref{lemma:KKMN}, $\set{\delta\seq{u}(t)\bmod p:
\seq{u}(t)\equiv a\bmod p}\supset\uGF{p}$. Thus, in this case, $D_{{a}}
=\GF{p}\setminus\set{r_0{a}\bmod p}$. Now  we have $\pair{a}{\gamma
a+jp^\ell}\in\brlab[\ell+1]$
 for $j\in\Int\setminus(r_0a+p\Int)$.
Assign any $a'\in\Int$. Since $p\geq3$, we can choose $j_0\in\Int$
satisfying $j_0 \not\equiv r_0{a}\bmod p$ and $j_0 \not\equiv
a'/{\gamma}-r_0{a}\bmod p$.
Notice that
\begin{equation*}
\gamma^2 a + a' p^\ell \not\equiv \gamma^2 a + \gamma(j_0 + r_0 a)p^\ell
\equiv \gamma(\gamma a + j_0 p^\ell) + r_0 (\gamma a + j_0 p^\ell)p^\ell
\bmod p^{\ell+1}.
\end{equation*}
Thus, we have $\pair{a}{\gamma a+j_0
p^\ell}\in\brlab[\ell+1]$ and $\pair{\gamma a + j_0 p^\ell}{\gamma^2a +
a' p^\ell} \in\brlab[\ell+1]$.
\end{enumerate}

Fix any $a\in \ggset$ and $b = a + a' p^\ell \in\pring$.
Let $k = \min\set{i>0: \gamma^i\equiv1\bmod p^{\ell}}$.
Denote $a_0=a$ and $a_k=b$.
For Cases (i) and (ii), we choose $a_j = a\gamma^j$
and have $\pair{a_{j}}{a_{j+1}}\in\brlab[\ell+1]$, $0\leq j< k$.
For Case (iii), we iteratively choose $a_j\in a\gamma^j + p^\ell\pring$
satisfying $a_j \not\equiv  a_{j-1}\gamma + r_0a_{j-1}p^\ell\bmod
p^{\ell+1}$, $1\leq j \leq k-2$, and $a_{k-1}\in a\gamma^{k-1} +
p^\ell\pring$ satisfying
\begin{equation*}
    \left\{
    \begin{array}{ccl}
a_{k-1} &\not\equiv&
a_{k-2}\gamma + r_0a_{k-2} p^\ell\bmod p^{\ell+1},\\
a_{k-1} &\not\equiv& a_{k-2}\gamma +
(a'/\gamma-r_0a_{k-2}) p^\ell\bmod p^{\ell+1}.
     \end{array}\right.
\end{equation*}
Then $\pair{a_{j-1}}{a_{j}}\in\brlab[\ell+1]$, $1\leq j\leq k$.
Therefore,  Statement \ref{stat:strong-trans-rt} holds for $l=\ell+1$.\fp

\begin{lemma}\label{lemma:rtl-p-3}
Let $\seq{u}$, $\seq{v}$, $\ell$ and
$\gamma$ be defined as above, and $1\leq\ell<\tail<e$.
If $\prmpoly$ is primitive and Statement \ref{stat:strong-trans-rt} holds for $l=\tail$, then
 Statement \ref{stat:strong-trans-rt} also holds for $l=\tail + 1$.
\end{lemma}
\pf Assign any $a\in \ggset$ and $d\in\Int 
$. Let $b = a + dp^{\tail}$.

Since Statement \ref{stat:strong-trans-rt} holds for $l=\tail$,
there exists $a_0,a_1,\dots,a_k$ such that
\begin{equation*}
    \left\{
     \begin{array}{l}
       a_0 \equiv a\mod p^\tail, \\
       a_k \equiv a+dp^{\tail-1}\mod p^\tail, \\
       \gamma^k\equiv1\mod p^\ell,\\
       \pair{a_{j-1}}{a_j}\in\brlab[\tail], 1\leq j\leq k.
     \end{array}\right.
\end{equation*}
As defined in $\brlab[\tail]$, there exist
$t_0,t_1,t_2,\dots,t_{k-1}\in\Int$ such that
\begin{equation*}
    \left\{
     \begin{array}{lcl}
       \delta\seq{u}(t_i)&\not\equiv& 0\mod p, 0\leq i< k,\\
       \seq{u}(t_i) & \equiv & \seq{v}(t_{i-1}) \equiv a_i \mod p^\tail, 1\leq i< k, \\
       \seq{u}(t_0) & \equiv & a \mod p^\tail, \\
       \seq{v}(t_{k-1}) & \equiv & a+dp^{\tail-1} \mod p^\tail.
     \end{array}\right.
\end{equation*}
For $1\leq i< k$, we iteratively define
\begin{equation*}
    \left\{
     \begin{array}{ccl}
       t_{0,0} &=& t_{0} + \perb{p^{\tail-1}d_{0,0}},\\
       t_{i,0} &=& t_{i} + \perb{p^{\tail-1}d_{i,0}},
     \end{array}\right.
\end{equation*}
where
\begin{equation*}
    \left\{
     \begin{array}{ccl}
        d_{0,0} &\equiv & \left(a-\seq{u}(t_{0})\right)/
\left(p^\tail\delta\seq{u}(t_{0})\right)
\mod p,\\
        d_{i,0} &\equiv & \left(\seq{v}(t_{i-1,0})-\seq{u}({t_{i}})\right)
/\left(p^\tail\delta\seq{u}(t_{i})\right) \mod p.
     \end{array}\right.
\end{equation*}
Since $\seq{u}(t_{0})\equiv a\bmod p^\tail$, $d_{0,0}$ is
well-defined. By Lemma \ref{lemma:seq-p-adic},
$\seq{v}(t_{i-1,0})\equiv\seq{v}(t_{i-1}) \equiv\seq{u}(t_{i}) \bmod
p^\tail$, $1\leq i<k$. Hence, $d_{i,0}$ is well-defined, $1\leq i<k$.
By Lemma \ref{lemma:seq-p-adic},
for $1\leq i<k$, we check that
\begin{equation}\label{eqn:conn-rt-t0}
    \left\{
     \begin{array}{cclcl}
       \seq{u}(t_{0,0}) & \equiv & \seq{u}(t_0) + p^\tail d_{0,0}\delta\seq{u}(t_0)  &\equiv &a \mod p^{\tail+1},\\
       \seq{u}(t_{i,0}) & \equiv & \seq{u}(t_i) + p^\tail d_{i,0}\delta\seq{u}(t_i)  &\equiv &\seq{v}(t_{i-1,0}) \mod p^{\tail+1}.
     \end{array}\right.
\end{equation}

Denote $\Delta=\seq{v}({t_{k-1,0}})-\seq{u}({t_{0,0}})$. Because
$\seq{v}(t_{k-1,0}) \equiv \seq{v}(t_{k-1}) \equiv a+dp^{\tail-1} \bmod p^\tail$
and $\seq{u}({t_{0,0}})\equiv
a\bmod p^\tail$, we have $\Delta\equiv dp^{\tail-1}\mod p^{\tail}$.

We choose 
$r_i=\gamma^i\Delta/
\left(p^{\tail-1}\delta\seq{u}(t_{i,0})\right)\bmod p$, $0\leq i
<k$. Let $t_{k,0} = t_{0,0} + \perb({r_0p^{\tail-2} +
p^{\tail-1}d_{k,0}})$, where $d_{k,0} \equiv
\left(\seq{v}(t_{k-1,0}) - \seq{u}(t_{0,0} + \perb r_0p^{\tail-2})\right)/
(p^\tail\delta\seq{u}(t_{0,0})) \bmod p$.
Then let $r_k = \gamma^k\Delta/
\left(p^{\tail-1}\delta\seq{u}(t_{k,0})\right)\bmod p$.

By Lemma \ref{lemma:prim-seq-mod-p}, $\delta\seq{u}\bmod p$ is an
$m$-sequence generated by $\prmpoly\bmod p$, and hence
$\delta\seq{u}(t_{i,0})\equiv\delta\seq{u}(t_i)\not\equiv0\bmod p$,
$0\leq i<k$.
Moreover, by Lemma
\ref{lemma:seq-p-adic}, $\seq{u}(t_{0,0} +
r_0p^{\tail-2}\perb) \equiv \seq{u}(t_{0,0}) +
r_0p^{\tail-1}\delta\seq{u}(t_{0,0}) \equiv \seq{v}(t_{k-1,0})\bmod
p^\tail$. Hence, $d_{k,0}$ is well-defined. By Lemma
\ref{lemma:seq-p-adic}, we check that
\begin{align}
\seq{u}(t_{k,0})\equiv & \seq{u}((t_{0,0} + \perb r_0p^{\tail-2}) +\perb
p^{\tail-1}d_{k,0})\notag \\
\equiv&
\seq{u}(t_{0,0} + \perb r_0p^{\tail-2}) +
p^\tail d_{k,0}\delta\seq{u}(t_{0,0} + \perb r_0p^{\tail-2})\notag\\
\equiv &
\seq{u}(t_{0,0} + \perb r_0p^{\tail-2}) +
p^\tail d_{k,0}\delta\seq{u}(t_{0,0})\notag\\
\equiv &
\seq{v}(t_{k-1,0})\bmod p^{\tail+1}.\label{eqn:conn-rt-t0k}
\end{align}
Similarly, by Lemma \ref{lemma:prim-seq-mod-p},
$\delta\seq{u}(t_{k,0})\equiv\delta\seq{u}(t_{0,0})\not\equiv0\bmod p$.
Besides, $\gamma^k\equiv 1\bmod p^\ell$.
Thus, we have $r_0\equiv r_k\bmod p$.

For $1\leq i\leq k$ and $1\leq j<p$, we iteratively define
\begin{equation}\label{eqn:rt-iter-def}
\left\{
    \begin{array}{ccl}
       {t_{0,j}} &=& {t_{k,j-1}}, \\ 
       t_{i,j}   &=&t_{i,0} + (p^\len-1){(jr_i + d_{i,j}p)p^{\tail-2}},
     \end{array}
\right.
\end{equation}
where
\begin{equation*}
d_{i,j}\equiv \left(\seq{v}({t_{i-1,j}}) -
\seq{u}\left({t_{i,0}+(p^\len-1){jr_ip^{\tail-2}}}\right)\right)
/\left(p^\tail\delta\seq{u}({t_{i,0}})\right)\bmod p.
\end{equation*}
For $1\leq j\leq p-1$, considering $r_0\equiv r_k\bmod p$, we
compute \allowdisplaybreaks
\begin{align}
t_{0,j}\equiv & t_{k,j-1} \equiv t_{k,0}+(p^\len-1){(j-1)r_k p^{\tail-2}} \notag\\
\equiv &t_{0,0}+(p^\len-1){(r_0+(j-1)r_k)p^{\tail-2}}\notag\\
\equiv &t_{0,0}+(p^\len-1){jr_0p^{\tail-2}}\bmod
p^{\tail-1}.\label{eqn:rt-iter-0}
\end{align}
For $1\leq i\leq k$ and $1\leq j\leq p-1$,
considering Eq.(\ref{eqn:conn-rt-t0}), Eq.(\ref{eqn:conn-rt-t0k}) and
$\seq{v}\equiv \gamma\seq{u}\bmod p$,
we use Eq.(\ref{eqn:seq-shift-funda})
to compute
\allowdisplaybreaks
\begin{align*}
      \seq{v}({t_{i-1,j}})
\equiv & \seq{v}\left({t_{i-1,0} + (p^\len-1){jr_{i-1}p^{\tail-2}}}\right)\\
\equiv & \seq{v}({t_{i-1,0}}) + jr_{i-1}p^{\tail-1}
 \delta\seq{v}({t_{i-1,0}})\\
 \equiv & \seq{v}({t_{i-1,0}}) + jr_{i-1}\gamma p^{\tail-1}
 \delta\seq{u}({t_{i-1,0}})\\
\equiv &\seq{u}({t_{i,0}}) + j\gamma^{i}\Delta\\
\equiv &\seq{u}({t_{i,0}}) + j r_i p^{\tail-1}\delta\seq{u}({t_{i,0}})\\
\equiv &\seq{u}\left({t_{i,0} + (p^\len-1){jr_ip^{\tail-2}}}\right)
 \mod p^\tail.
\end{align*}
Therefore, $d_{i,j}$s are well-defined,
and moreover, we can use Lemma \ref{lemma:seq-p-adic} to check
\begin{align}
\seq{u}(t_{i,j})\equiv &
\seq{u}(t_{i,0} + \perb{(jr_i + d_{i,j}p)p^{\tail-2}})\notag \\
\equiv&
\seq{u}((t_{i,0} + \perb jr_ip^{\tail-2}) +
\perb p^{\tail-1} d_{i,j})\notag\\
\equiv&
\seq{u}(t_{i,0} + \perb jr_ip^{\tail-2})
+ p^\tail d_{i,j}\delta\seq{u}(t_{i,0} + \perb jr_ip^{\tail-2})
\notag\\
\equiv &
\seq{u}(t_{i,0} + \perb jr_ip^{\tail-2})
+ p^\tail d_{i,j}\delta\seq{u}(t_{i,0})\notag\\
\equiv &
\seq{v}(t_{i-1,j})\bmod p^{\tail+1}\label{eqn:conn-rt-tij}
\end{align}
for $1\leq i\leq k$, $1\leq j\leq p-1$.
Notice
$\delta\seq{u}(t_{i-1,j})\equiv\delta\seq{u}(t_{i-1,0})
\not\equiv0\bmod p$
and $\pair{\seq{u}({t_{i-1,j}})}{\seq{v}({t_{i-1,j}})}\in\brlab$.
Thus, combining Eq.(\ref{eqn:conn-rt-t0}),
Eq.(\ref{eqn:conn-rt-t0k}) and Eq.(\ref{eqn:conn-rt-tij}),
we have
\begin{equation}\label{eqn:conn-rt-pf-all}
\pair{\seq{u}({t_{i-1,j}})}{\seq{u}({t_{i,j}})}\in\brlab[\tail+1],
1\leq i\leq k,0\leq j<p.
\end{equation}

Denote $\seq{w}=(\seq{v}-\gamma\seq{u})/p^\ell$.
We apply Lemma \ref{lemma:seq-p-adic} on $p^\ell\seq{w}$
to get $\seq{w}(t_{i,j})-\seq{w}(t_{i,0})\equiv jr_ip^{\tail-1}\delta\seq{w}(t_{i,0})\bmod p^\tail$.
Then for $0\leq i< k$, we have
\allowdisplaybreaks
\begin{align}
      &\seq{u}({t_{i+1,j}})-\seq{u}({t_{i+1,0}})
      \equiv\seq{v}{(t_{i,j})} - \seq{v}(t_{i,0}) \notag \\
\equiv&\gamma(\seq{u}({t_{i,j}})-\seq{u}({t_{i,0}})) +
p^{\ell}(\seq{w}(t_{i,j})-\seq{w}(t_{i,0}))\notag \\
\equiv&\gamma(\seq{u}({t_{i,j}})-\seq{u}({t_{i,0}})) +
p^{\ell+\tail-1}jr_i \delta\seq{w}({t_{i,0}})\mod
p^{\tail+1}.\label{eqn:conn-rt-iter}
\end{align}
Thus, from Lemma \ref{lemma:seq-p-adic},
Eq.(\ref{eqn:rt-iter-0}),
Eq.(\ref{eqn:conn-rt-iter})
and $\gamma^k\equiv1\bmod p^\ell$,
we have
\allowdisplaybreaks
\begin{align*}
 &(\seq{u}({t_{k,j}})-\seq{u}({t_{0,j}}))-(\seq{u}({t_{k,0}})-\seq{u}({t_{0,0}}))\\
\equiv&(\seq{u}({t_{k,j}})-\seq{u}({t_{k,0}}))-(\seq{u}({t_{0,j}}) - \seq{u}({t_{0,0}})) \\
\equiv&(\gamma^k-1)(\seq{u}({t_{0,j}})-\seq{u}({t_{0,0}}))
 + j p^{\ell+\tail-1} \sum_{i=0}^{k-1} \gamma^{k-1-i}r_i
 \delta\seq{w}({t_{i,0}})\\
\equiv&p^{\ell+\tail-1}jr_0\delta\seq{u}({t_{0,0}})
 + j p^{\ell+\tail-1} \sum_{i=0}^{k-1} \gamma^{k-1-i}r_i
 \delta\seq{w}({t_{i,0}})
 \mod p^{\tail+1}
\end{align*}
for $1\leq j\leq p-1$. Denoting
$\Delta'=\seq{u}({t_{k,p-1}})-\seq{u}({t_{0,0}})$, we compute
 \allowdisplaybreaks
\begin{align*}
\Delta'
\equiv & \sum_{j=0}^{p-1}(\seq{u}({t_{k,j}}) - \seq{u}({t_{0,j}}))\\
\equiv & p\Delta +
p^{\ell+\tail-1}r_0\delta\seq{u}({t_{0,0}})\sum_{j=1}^{p-1}j  \\ &
 +
 p^{\ell+\tail-1} \sum_{i=0}^{k-1}\gamma^{k-1-i}r_i \delta\seq{w}({t_{i,0}})
 \sum_{j=1}^{p-1}j  \mod p^{\tail+1}.
\end{align*}
Since $\sum_{j=1}^{p-1}j \equiv p(p-1)/2\equiv 0\bmod p$,
we have $\Delta'\equiv p\Delta \equiv dp^{\tail}\mod p^{\tail+1}$.
Thus, we have
\begin{equation}\label{eqn:conn-rt-tkp}
\seq{u}(t_{k,p-1})\equiv
\seq{u}(t_{0,0})+\Delta' \equiv a + dp^\tail\equiv b\mod p^{\tail+1}.
\end{equation}

Let $a'_{pk}=\seq{u}(t_{k,p-1})$ and
$a'_{i+kj} = \seq{u}(t_{i,j})$, $0\leq i<k$, $0\leq j<p$.
Then we have $a'_0,a'_1\dots,a'_{pk}$ satisfying
\begin{equation*}
    \left\{
     \begin{array}{ll}
       a'_0 \equiv \seq{u}(t_{0,0})\equiv a\mod p^{\tail+1},
       & \text{by Eq.(\ref{eqn:conn-rt-t0})} \\
       a'_{pk} \equiv\seq{u}(t_{k,p-1})\equiv b\mod p^{\tail+1},
       &       \text{by Eq.(\ref{eqn:conn-rt-tkp}})\\
       \gamma^{pk}\equiv(\gamma^k)^p\equiv1\mod p^\ell,&\\
       \pair{a'_{j-1}}{a'_j}\in\brlab[\tail+1], 1\leq j\leq pk.& \text{by Eq.(\ref{eqn:rt-iter-def}) and Eq.(\ref{eqn:conn-rt-pf-all})}
     \end{array}\right.
\end{equation*}
Since $\tail\geq\ell+1$,
$\pair{a'_{j-1}}{a'_j}\in\brlab[\tail+1]$ ensures
that $a'_j\in\set{a\gamma^t:t\in\Int}+p^\ell\pring$.
Therefore, Statement \ref{stat:strong-trans-rt} holds
for $l=\tail+1$. \fp

Now we combine Lemma \ref{lemma:local-branch-irrt}
and Statement \ref{stat:strong-trans-rt} to
describe the equivalence classes w.r.t. $\rlabe$.
\begin{lemma}\label{lemma:conn-rt-equiv-clss}
Let $\seq{u}$, $\seq{v}$, $\ell$ and $\gamma$ be defined as above,
and $1\leq\ell\leq e$.
If $\prmpoly$ is strongly primitive, then
$\set{\set{a\gamma^i:i\in\Int} + p^{\ell}\pring: a\in\pring}$
is the set of equivalence classes w.r.t. $\rlabe$.
If $\prmpoly$ is primitive but not strongly primitive, then
for any $ a\in\gset$,
$\set{a\gamma^i:i\in\Int} + p^{\ell}\pring$
is an equivalence class w.r.t. $\rlabe$.
\end{lemma}
\pf  Notice $\brlab[i]\subset\rlab[i]\subset\rlabs[i]$,
$1\leq i\leq e$.
For $a\in\pring$, denote
$C_a = \set{a\gamma^i:i\in\Int} + p^{\ell}\pring$.

If $\ell=e$, $\rlab=\set{\pair{r}{\gamma r}:r\in\set{\seq{u}(t):t\in\Int}}$.
Since $\gset$ is a finite cyclic group,
the equivalence class of $a\in \ggset$ w.r.t. $\rlabe$ is
$C_a=\set{a\gamma^i:i\in\Int}$.

Now we consider $1\leq \ell<e$.

First, we show that for $a \in\ggset$, $C_a$ is a subset of an equivalence class w.r.t. $\rlabe$.
Fix any $a\in \ggset$ and $b = a \gamma^k + b' p^\ell  \in C_a$.
There exist $t_0\in\Int$ satisfying $\seq{u}(t_0)=a$.
Iteratively, for $1\leq i\leq k$,
because $\seq{v}(t_{i-1})\equiv \gamma \seq{u}(t_{i-1})\equiv a\gamma^{i}\bmod p^\ell$,
we have $\seq{v}(t_{i-1})\in C_a\subset \ggset\subset \set{\seq{u}(t):t\in\Int}$, 
and then there also exists
$t_i\in\Int$ satisfying $\seq{u}(t_i)=\seq{v}(t_{i-1})$.
Then $\pair{\seq{u}(t_{j-1})}{\seq{u}(t_j)}\in\rlab\subset\rlab[\ell+1]$,
$1\leq j\leq k$.
Assumption
(i) of Lemma \ref{lemma:local-branch-irrt} holds for $B=C_a$ and
$\tau=\ell$.
Furthermore, by Lemma \ref{lemma:rt-conn-ell} and Lemma
\ref{lemma:rtl-p-3}, we have inductively proved Statement \ref{stat:strong-trans-rt}
for $\ell<i\leq e$, i.e., for any $a\in\pring$ and
$b\in a+p^{i-1}\pring$, there exist $a_0,a_1\dots,a_k\in C_a$
satisfying $a_0\equiv a\bmod p^i$, $a_k\equiv b\bmod p^i$ and
$\pair{a_{j-1}}{a_j}\in\brlab[i]\subset\rlabs[i]$, $1\leq j\leq k$.
Thus, Assumption (ii) of Lemma
\ref{lemma:local-branch-irrt} holds for $B=C_a$ and $\tau=\ell$. By
Lemma \ref{lemma:local-branch-irrt}, $C_a\times C_a\subset\rlabe$.

Second, noticing $\seq{v}\equiv\gamma\seq{u}\bmod p^\ell$,
we conclude that
if $\pair{a}{b}\in\rlabe$, then $b\equiv a\gamma^i\bmod p^\ell$
for some $i\in\Int$, i.e., $b\in C_a$.

Thus, for $a\in\ggset$,
$C_a$ is an equivalence class w.r.t. $\rlabe$.

Now suppose $\prmpoly$ to be strongly primitive.
We have to prove that $\set{C_a:a\in\pring}$ is
the set of equivalence classes w.r.t. $\rlabe$.

On one hand, we show
that for $a,b\in\pring$, either $C_a=C_b$ or $C_a\cap
C_b=\emptyset$.
Notice that
\begin{equation*}
C_a=\set{r\in\pring:\exists i\in\Int, r\equiv
a\gamma^i\bmod p^\ell}.
\end{equation*}
Assume $a'\in C_a$ and $a'\equiv
a\gamma^{i_0}\bmod p^\ell$. Since $\gamma\not\equiv0\bmod p$, given
$r\in C_a$ satisfying $r\equiv a\gamma^i\bmod p^\ell$, we have
$r\equiv a\gamma^i\equiv a'\gamma^{i-i_0}\bmod p^\ell$. Hence,
$C_a\subset C_{a'}$. We see $a\in C_{a'}$ and similarly have $C_{a'}\subset
C_a$. Thus, $C_a=C_{a'}$ for any $a'\in C_a$.
 Suppose $C_a\cap C_b\neq\emptyset$, say $c\in
C_a\cap  C_b$. Then $C_a=C_c=C_b$ as shown above.

On the other hand, $\cup_{a\in\pring} C_a=\pring$. We conclude that
$\set{C_a: a\in\pring}$
is the set of equivalence classes w.r.t. $\rlabe$.
\fp

\subsection{Proof of main results}
\label{subsect:pf-main-thm}
In this subsection we prove Theorem \ref{thm:result-uniform},
Theorem \ref{thm:uniform-branch-p-3},
Theorem \ref{thm:uniform-result-hyper-prim-p3}
and Corollary \ref{cor:num-ep-map-p3}.

\pf[Proof of Theorem \ref{thm:result-uniform}]
If $\prmpoly$ is primitive and $(\delta^2\bmod(p,\sigma))\notin\GF{p}$,
we use Lemma \ref{lemma:equiv-class-irrt} and  Lemma
\ref{lemma:conn-rt-equiv-clss} to obtain $\mathscr{C}$ as the set
of equivalence classes w.r.t. $\rlabe$.
Then this theorem follows from Lemma \ref{lemma:map-eqrl}.\fp

We need the following lemma to describe three important kinds of
subset of equivalence classes w.r.t. $\rlabe$.
\begin{lemma}\label{lemma:equiv-cls-sub}
Let $\prmpoly$ be a primitive polynomial over $\pring$
and $\seq{u},\seq{v}\in\gseq$. If $\seq{u}\neq \seq{v}$,
then at least one of the following three statements is true:
(i) There exists a nontrivial $k$-th root of unity
$\gamma\in\pring$, 
such that
$\seq{v}=\gamma\seq{u}$ and
$\set{a\gamma^i:1\leq i\leq k}$ is an equivalence class w.r.t. $\rlabe$
for any $a\in\gset$ (for any $a\in{\pring}$ if $\prmpoly$ is strongly primitive).
(ii) For any $a\in \gset$, $a + p^{e-1}\pring$ is a subset
of an equivalence class w.r.t. $\rlabe$.
(iii) Condition \ref{cond:pit-p-lt-3} holds,
and for any $a\in p\pring$,
$a + p^{e-1}\pring$ is a subset
of an equivalence class w.r.t. $\rlabe$.
\end{lemma}
\pf
Let $\ell=\max\set{0\leq i\leq e:\exists j\in \Int, j\seq{u}\equiv
\seq{v}\bmod p^i}$ and let $\gamma\in\pring$
satisfy $\gamma\seq{u}\equiv\seq{v}\bmod p^\ell$.

Suppose $\ell=e$, i.e., there exists $\gamma\in\pring$ satisfying
$\seq{v}=\gamma\seq{u}$.
By Lemma \ref{lemma:conn-rt-equiv-clss}, for any $a\in\gset$
(for any $a\in\pring$ if $\prmpoly$ is strongly primitive),
$\set{a\gamma^t:t\in\Int}$ is an equivalence class w.r.t. $\rlabe$.
Consider two cases.
\begin{itemize}
    \item Case I: $\gamma^{p-1}\equiv1\bmod p^e$.
Let $k=\min\set{j>0:\gamma^j\equiv1\bmod p^e}$.
Then $\seq{v}=\gamma\seq{u}$ and
$\set{a\gamma^i:1\leq i \leq r}$ is an equivalence class w.r.t. $\rlabe$
for any $a\in\gset$ (for any $a\in\pring$ if $\prmpoly$ is strongly primitive).
Statement (i) of Lemma \ref{lemma:equiv-cls-sub} is true.
    \item Case II: $\gamma^{p-1}\not\equiv1\bmod p^e$.
Since $\gamma^{p-1}\equiv1\bmod p$, we have $\gamma^{p-1}\equiv1+\lambda p^i\bmod p^e$
for some $1\leq i<e$ and $\lambda\in\gset$.
Letting  $\lambda^{-1}\in\Int$ satisfy $\gamma\gamma^{-1}\equiv1\bmod p^e$,
we have $(1+\lambda p^i)^{\lambda^{-1}}\equiv 1 + \lambda \lambda^{-1}p^i\equiv 1+ p^i\bmod p^{i+1}$.
For any $a'\equiv1\bmod p$ and $j\geq 1$,
$(1+ a'p^j)^{p}\equiv1+a'p^{j+1}\bmod p^{j+2}$.
Inductively, we have
$\gamma^{(p-1)\lambda^{-1}p^{e-1-i}}\equiv1+ p^{e-1}\bmod p^e$.
Thus, $1+p^{e-1}\in\set{\gamma^t:t\in\Int}$.
For $a\in\gset$,
$a+p^{e-1}\pring = \set{a(1+p^e)^t:t\in\Int}
\subset\set{a\gamma^t:t\in\Int}$.
Thus, Statement (ii) of Lemma \ref{lemma:equiv-cls-sub} holds.
\end{itemize}

Suppose $1\leq \ell<e$.
By Lemma \ref{lemma:conn-rt-equiv-clss},
for any $a\in \gset$,
$a+p^{e-1}\pring\subset
\set{a\gamma^t:t\in\Int}+p^\ell\pring$
and hence is a subset of an equivalence class.
Thus, Statement (ii) of Lemma \ref{lemma:equiv-cls-sub} holds.

Suppose $\ell=0$, i.e., $\seq{u}\linind\seq{v}$.
Consider three cases.
\begin{itemize}
  \item Case I: $\prmpoly$ is strongly primitive and Condition \ref{cond:pit-p-lt-3} does
not hold.
By  Lemma \ref{lemma:equiv-class-irrt},
$\pring$ is the equivalence class.
Hence, the statement (i)-(iii) of Lemma \ref{lemma:equiv-cls-sub}
hold.
   \item Case II:  Condition \ref{cond:pit-p-lt-3} holds.
By Lemma \ref{lemma:conn-irt-cnd1},
$a+p^{e-1}\pring$ is a subset of an equivalence class
for $a\in p\pring$, and hence
Statement (iii) of Lemma \ref{lemma:equiv-cls-sub} is true.
  \item Case III: $\prmpoly$ is primitive but not strongly primitive.
By Lemma \ref{lemma:conn-irt-prm}, $a+p^{e-1}\pring$ is a subset of an equivalence class
for $a\in\gset$, and hence
Statement (ii) of Lemma \ref{lemma:equiv-cls-sub} is true.
\end{itemize}

We have listed all possible cases.
Therefore, at least one of the statements (i)-(iii)
of Lemma \ref{lemma:equiv-cls-sub}
is true.\fp

\pf[Proof of Theorem \ref{thm:uniform-branch-p-3}]
Let $\seq{u},\seq{v}\in\gseq$.
Since  $\seq{u}=\seq{v}$ implies  $D$-uniformity
of $\indmap{\cmap}(\seq{u})$ and $\indmap{\cmap}(\seq{v})$,
we only have to prove that under the three statements (i)-(iii)
of Theorem \ref{thm:uniform-branch-p-3},
if $\seq{u}\neq\seq{v}$,
then $\indmap{\cmap}(\seq{u})$ and
$\indmap{\cmap}(\seq{v})$ are not $D$-uniform.

Assume $\seq{u}\neq\seq{v}$.
By Lemma \ref{lemma:map-eqrl}, it is sufficient to show that
under the three statements (i)-(iii)
of Theorem \ref{thm:uniform-branch-p-3},
$\cmap$ is not constant on some equivalence class $C$ w.r.t. $\rlabe$,
where $\cmap(C)\cap D\neq\emptyset$.

By Lemma \ref{lemma:equiv-cls-sub},
at least one of the three statements (i)-(iii) of Lemma \ref{lemma:equiv-cls-sub}
is true. We consider the three possible scenarios.
\begin{itemize}
  \item
Scenario I: Statement (i) of Lemma \ref{lemma:equiv-cls-sub} holds.
Noticing $k\geq2$, we choose $r$ to be a prime
divisor of $k$ and $\omega=\gamma^{k/r}$.
Because of Statement (i) of Theorem \ref{thm:uniform-branch-p-3},
there exists $a\in\gset$ ($a\in\pring$ if $\prmpoly$ is strongly primitive)
such  that $\cmap(a)\in D$ and $\cmap$ is not constant on
$\set{a\omega^i:1\leq i \leq r}$.
Denote $C=\set{a\gamma^t:t\in\Int}$.
Then $C$ is the equivalence class of $a$ and $\cmap(C)\cap D\supset\set{\cmap(a)}$.
However, $\cmap$ is not constant $C$.
\item
Scenario II: Statement (ii) of Lemma \ref{lemma:equiv-cls-sub} holds.
Because of Statement (ii) of Theorem \ref{thm:uniform-branch-p-3},
there exists $a\in\gset$ such that  $\cmap(a)\in D$ and $\cmap$
is not constant on $a+p^{e-1}\pring$.
Let $C$ be the equivalence class of $a$ w.r.t. $\rlabe$,
then $a+p^{e-1}\pring\subset C$.
Thus, $\cmap(C)\cap D\supset\set{\cmap(a)}$
and $\cmap$ is not constant on $C$.
\item
Scenario III: Statement (iii) of Lemma \ref{lemma:equiv-cls-sub} holds.
Because of Statement (iii) of Theorem \ref{thm:uniform-branch-p-3},
there exists $a\in p\pring$ such that $\cmap(a)\in D$ and $\cmap$
is not constant on $a+p^{e-1}\pring$.
Let $C$ be the equivalence class of $a$ w.r.t. $\rlabe$,
then $a+p^{e-1}\pring\subset C$.
Thus, $\cmap(C)\cap D\supset\set{\cmap(a)}$
and $\cmap$ is not constant on $C$.
\end{itemize}
Therefore, if the three statements (i)-(iii)
of Theorem \ref{thm:uniform-branch-p-3} hold,
then $\cmap$ is not constant on some equivalence class $C$ w.r.t. $\rlabe$,
where $\cmap(C)\cap D\neq\emptyset$,
and hence $\indmap{\cmap}(\seq{u})$ and $\indmap{\cmap}(\seq{v})$
are not $D$-uniform.
\fp

\pf[Proof of Theorem \ref{thm:uniform-result-hyper-prim-p3}] Because
$({\delta}^2\bmod (p,\prmpoly))\notin\GF{p}$, $\prmpoly$ is strongly
primitive and Condition \ref{cond:pit-p-lt-3} does not hold.

Noticing that Statement (iii) of Lemma \ref{lemma:equiv-cls-sub}
does not occur here,
we can repeat the proof of Theorem \ref{thm:uniform-branch-p-3}
to show that
if Statement (i) and Statement (ii) of Theorem \ref{thm:uniform-result-hyper-prim-p3}
hold, then
$\indmap{\cmap}$ is injective on $\gseq$ w.r.t. $D$-uniformity.

Now suppose
either Statement (i) or Statement (ii) of Theorem \ref{thm:uniform-result-hyper-prim-p3}
does not hold.
Below we show that $\indmap{\cmap}$ is not injective on $\gseq$ w.r.t. $D$-uniformity.

If Statement (i) of Theorem \ref{thm:uniform-result-hyper-prim-p3} is not true, then there exists
a nontrivial $r$-th root of unity $\omega\in\pring$, such that
for any $a\in\pring$ with $\cmap(a)\in D$, $\cmap$
is constant on $\set{a\omega^i:1\leq i\leq r}$.
We take $\seq{u}\in\gseq$ and $\seq{v}=\omega\seq{u}$. Then
$\seq{v}\in\gseq$, and for any $d\in D$,
$\indmap{\cmap}(\seq{u})(t)=\cmap(\seq{u}(t)) = d$
 if and only if
$\indmap{\cmap}(\seq{v})(t)=\cmap(\seq{v}(t))
 = \cmap(\omega\seq{u}(t)) = d$.
Thus, we have $\seq{u}\neq\seq{v}$,
but $\indmap{\cmap}(\seq{u})$ and  $\indmap{\cmap}(\seq{v})$ are
$D$-uniform.

If Statement (ii) of Theorem \ref{thm:uniform-result-hyper-prim-p3} is not true, then for any $a\in\gset$
with $\cmap(a)\in D$,
$\cmap$ is constant on $a+p^{e-1}\pring$.
We take $\seq{u}\in\gseq$ and
$\seq{v}=(1+p^{e-1})\seq{u}$.
Then $\seq{v}\in\gseq$.
Since
\begin{equation*}
(1+p^{e-1})\seq{u}(t)=\seq{u}(t)+p^{e-1}\seq{u}(t)
\left\{
\begin{array}{l@{\text{ if }}l}
   = \seq{u}(t), & \seq{u}(t)\in p\pring,\\
   \in\seq{u}(t)+p^{e-1}\pring, & \seq{u}(t)\in \gset,
\end{array}\right.
\end{equation*}
we conclude that for any $d\in D$,
$\indmap{\cmap}(\seq{u})(t)=\cmap(\seq{u}(t)) = d$
 if and only if
$\indmap{\cmap}(\seq{v})(t)=\cmap(\seq{v}(t))
 = \cmap((1+p^{e-1})\seq{u}(t)) = d$.
Thus, we have $\seq{u}\neq\seq{v}$,
but $\indmap{\cmap}(\seq{u})$ and  $\indmap{\cmap}(\seq{v})$ are
$D$-uniform.
\fp

\pf[Proof of Corollary \ref{cor:num-ep-map-p3}] A non-entropy-preserving map $\cmap:\pring\rightarrow \imset$
satisfies neither Statement (i) nor Statement (ii) of Corollary
\ref{thm:result-hyper-prim-p3}. If $\cmap$ does not satisfy
Statement (i) of Corollary \ref{thm:result-hyper-prim-p3},
then there exists a prime divisor $r$ of $p-1$ such that $\cmap$ is constant on
$\set{a \omega^i: 1\leq i\leq r}$ for any $a\in\pring$, where
$\omega$ is a nontrivial root of unity satisfying $\omega^r=1$.
Since $\cset{\set{\set{a \omega^i: 1\leq i\leq r}:a\in\pring}} =
(p^e-1)/r + 1$, a map disobeying Statement (i) of Corollary
\ref{thm:result-hyper-prim-p3} is determined by the images of
$(p^e-1)/r + 1$ elements. Hence, the number of maps disobeying Statement (i)
of Corollary \ref{thm:result-hyper-prim-p3} is $k^{1+(p^e-1)/r}$. If
$\cmap$ does not satisfy Statement (ii) of Corollary
\ref{thm:result-hyper-prim-p3}, then $\cmap$ is constant on $a +
p^{e-1}\pring$ for $a\in\gset$. A map disobeying Statement (ii) of
Corollary \ref{thm:result-hyper-prim-p3}
is determined by the images of $(p^e-p^{e-1})/p + p^{e-1}$ elements.
Hence, the number of
maps disobeying Statement (ii) of Corollary
\ref{thm:result-hyper-prim-p3} is
$k^{p^{e-1}+(p^e-p^{e-1})/p}=k^{2p^{e-1}-p^{e-2}}$. Moreover,
constant maps satisfy neither Statement (i) nor Statement (ii) of
Corollary \ref{thm:result-hyper-prim-p3}, and are counted in both
cases above. Therefore, the number of entropy-preserving maps from
$\pring$ to $\imset$ is greater than \allowdisplaybreaks
\begin{align*}
      & k^{p^e}-k^{2p^{e-1}-p^{e-2}}-\sum_{i\in P}k^{1+(p^e-1)/r}\\
    > & k^{p^e}-k^{2p^{e-1}-p^{e-2}}- k^{1+(p^e-1)/2}\log_2 p\\
    = & k^{p^e}\left(1-k^{-p^{e-2}(p-1)^2}- k^{(1-p^e)/2}\log_2 p
    \right),
\end{align*}
where $P=\set{j\geq2:j \text{ is prime},j\mid(p-1)}$. \fp

\subsection{Some specific compressing maps}
\label{subsect:new-map}
In this subsection we discuss
some specific compressing maps from $\pring$ to
$\GF{p}$, and prove Theorem \ref{thm:special-lin-ind},
Theorem \ref{thm:new-map-str},
Theorem \ref{thm:new-map-weak} and
Theorem \ref{thm:new-map-weak-sq}.

In this subsection
$a\in\pring$ is identified as its unique
representative in $\set{0,1,\dots,p^e-1}$,
and $\GF{p}$ is considered as the set $\set{0,1,\dots,p-1}$.
The $i$-th coordinate $\crd{a}\in\GF{p}$ of $a\in\pring$ is defined by
$\rep{a}=\crd[0]{a} + \crd[1]{a} p +\dots+\crd[e-1]{a}p^{e-1}$,
where $\crd{a}\in \GF{p}$.
For simplicity we
write $a_i$ instead of $\crd{a}$ where ambiguity is impossible.
For $1\leq i<e$ and $a,c\in\pring$, let $\cut[i]{a}$ denote the
integer satisfying $\cut[i]{a}\equiv a\bmod p^i$ and
$0\leq\cut[i]{a}<p^i$;
let $\crr[c]{i}{a}\in\GF{p}$ satisfy
$\crr[c]{i}{a}\equiv(c \cut[i]{a}-\cut[i]{c a})/p^i\bmod p$.

Following the convention of previous papers
\cite{JL14,JL14p,SQ,TQ07,ZxQ10,ZxQ13uni,ZQ04,ZQ07it,ZQ08},
$a$ is also identified with the vector
$(a_0,a_1,\dots,a_{e-1})\in\GF{p}^e$ and thereby a map
from $\pring$ to $\GF{p}$
is explicitly written as an $e$-variable function on $\GF{p}$.
Moreover, each function from $\GF{p}^e$ to $\GF{p}$ is written
as a multivariate polynomial
in which the degree in each indeterminate 
is less than $p$.

The following four lemmas are useful in our proof.

\begin{lemma}\label{lemma:carry-def-prty}
Let $\gamma,a\in\pring$ and $1\leq i<e$.
Then $\crd{\gamma a} \equiv
\crr[\gamma]{i}{a} + {\gamma}{a}_i\bmod p$. Particularly, for
$b\in\GF{p}$,
\begin{equation}\label{eqn:cmp-map-cry-gnl}
    \crd[e-1]{\gamma\left(a+p^{e-1}b\right)} \equiv \gamma(a_{e-1}+b)+\crr[\gamma]{e-1}{a}\mod p.
\end{equation}
\end{lemma}
\pf On one hand, $a \equiv \cut[i]{a} + p^i {a}_i\bmod p^{i+1}$ and
$\gamma a \equiv \cut[i]{\gamma a} + p^i \crd{\gamma a}\bmod
p^{i+1}$. On the other hand, from the definition of $\crr[\gamma]{i}{a}$
we have $ \gamma\cut[i]{a} \equiv
\cut[i]{\gamma a} + p^i \crr[\gamma]{i}{a}\bmod p^{i+1}$. Thus,
\begin{equation*}
    \cut[i]{\gamma a} + p^i \crd{\gamma a} \equiv
    \gamma a \equiv
    \gamma(\cut[i]{a} + p^i {a}_i)
    \equiv \cut[i]{\gamma a} + p^i \crr[\gamma]{i}{a} + p^i \gamma {a}_i
    \bmod p^{i+1},
\end{equation*}
implying $\crd{\gamma a} \equiv \crr[\gamma]{i}{a} +
{\gamma}{a}_i\bmod p$.
Particularly, $ \crd[e-1]{\gamma\left(a+p^{e-1}b\right)}
\equiv \gamma(a_{e-1}+b)+\crr[\gamma]{e-1}{\left(a+p^{e-1}b\right)}\bmod p$.
Furthermore, notice that
 $\crr[\gamma]{e-1}{a}=\crr[\gamma]{e-1}{a+jp^{e-1}}$ for any $j\in\Int$.
Therefore, Eq.(\ref{eqn:cmp-map-cry-gnl}) is true.\fp

\begin{lemma}\label{lemma:carry-sum-0}
Let $1\leq i<e$. Suppose  $1\neq\gamma\in\pring$
and $\gamma^r\equiv1\bmod p^e$, where $r>1$ and $r\mid(p-1)$.
If $p^i\nmid a$,
then there exists $k\in\set{0,2,\dots,r-1}$ satisfying
$\crr[\gamma]{i}{\gamma^k a}\neq0$.
\end{lemma}
\pf Use reductio ad absurdum. Suppose $\crr[\gamma]{i}{\gamma^j a}=0$ for
any $j\in\set{0,1,\dots,r-1}$. Then by Lemma
\ref{lemma:carry-def-prty}, we iteratively get $\crd{\gamma^{j+1} a}
\equiv
 \crr[\gamma]{i}{\gamma^{j}a}+{\gamma}\crd{\gamma^{j}
a} \equiv {\gamma}^{j+1} a_{i}\bmod p$, $0\leq j<r$. Since $1\leq\cut[i]{\gamma^j
a}<p^{i}$, we have $r\leq\sum_{i=0}^{r-1}\cut[i]{\gamma^ja}<rp^{i}$ and
hence $p^{i+1}\nmid\sum_{i=0}^{r-1}\cut[i]{\gamma^ja}$. However, seeing
$\rep{\gamma^j a}\equiv\cut[i]{\gamma^j a}+p^{i}\crd{\gamma^{j}a}
\bmod p^{i+1}$, we have \allowdisplaybreaks
\begin{align*}
\sum_{j=0}^{r-1}\cut[i]{\gamma^ja} \equiv &
 \sum_{j=0}^{r-1}\cut[i]{\gamma^ja}
+p^{i}a_{i}\sum_{j=0}^{r-1}{\gamma}^j\\
\equiv&\sum_{j=0}^{r-1}\cut[i]{\gamma^ja}
+p^{i}\sum_{j=0}^{r-1}\crd{\gamma^ja}\\
 \equiv&\sum_{j=0}^{r-1} \rep{\gamma^ja}
\equiv 0 \mod p^{i+1},
\end{align*}
contradictory to $p^{i+1}\nmid\sum_{i=0}^{r-1}\cut[i]{\gamma^ja}$.
Therefore, our supposition is ridiculous and
$\crr[\gamma]{i}{\gamma^k a}\neq0$ for some $0\leq k<r$.\fp

\begin{lemma}\label{lemma:carry-diff}
Let $1\leq i<e$. Suppose  $\gamma\in\pring$
and $\gamma^r\equiv1\bmod p^e$, where $r>1$ and $r\mid(p-1)$.
If $\gamma\not\equiv\pm1\bmod p^i$,
then there exist $a,b\in\set{0,1,\dots,p^{e-1}-1}$
satisfying $p\nmid a$, $p\nmid b$ and
$\crr[\gamma]{i}{a}\neq\crr[\gamma]{i}{b}$.
\end{lemma}
\pf
Let $a_1 = 1$, $a_2 = 2 $ and $a_3 = p^{i}-1$.
Then $p\nmid a_j$ and $\crd[i]{a_j}=0$,
$j\in\set{1,2,3}$.
By Lemma \ref{lemma:carry-def-prty},
we have $\crr[\gamma]{i}{a_j} = \crd[i]{\gamma a_j}$.

Clearly, $\crr[\gamma]{i}{\gamma a_1}  = \crd[i]{\gamma}$.
Noticing
\begin{equation*}
2\gamma \equiv 2(\cut[i]{\gamma}+\gamma_ip^i)\equiv 2\cut[i]{\gamma} + 2\gamma_ip^i\bmod p^{i+1}
\end{equation*}
and $\cut[i]{\gamma}<p^{i}$,
we compute
\begin{equation*}
\crr[\gamma]{i}{\gamma a_2}  = \crd[i]{2\gamma}
\equiv\left\{
\begin{array}{l@{\text{ if }}l}
  2 {\gamma}_i  \bmod p,   &  2\cut[i]{\gamma}<p^{i} ,\\
  2 {\gamma}_i + 1 \bmod p,&  2\cut[i]{\gamma}>p^{i}.
\end{array}\right.
\end{equation*}
Noticing $\crd[i]{-\gamma}=\crd[i]{ p^{e} - \gamma } = p-1-{\gamma}_i$,
we compute
\begin{equation*}
\crr[\gamma]{i}{\gamma a_3} = \crd[i]{\gamma (p^{i}-1)}
\equiv \gamma + \crd[i]{-\gamma}
\equiv \gamma + p - 1 -{\gamma}_i\bmod p.
\end{equation*}

We use reductio ad absurdum.
Assume $\crr[\gamma]{i}{a_1}=\crr[\gamma]{i}{a_2}=\crr[\gamma]{i}{a_3}$.
From $\crr[\gamma]{i}{a_1}=\crr[\gamma]{i}{a_3}$ we get
$\gamma_i \equiv \gamma + p -1 -\gamma_i\bmod p$,
implying $\gamma_i \equiv (\gamma - 1)/2 \bmod p$.
Consider two cases.
\begin{enumerate}
  \item[(i)] $2\cut[i]{\gamma} < p^{i}$.
From $\crr[\gamma]{i}{a_1}=\crr[\gamma]{i}{a_2}$ we get
$\gamma_i\equiv 2\gamma_i\equiv (\gamma-1)/2\bmod p$,
implying $\gamma_i=0$ and $\gamma\equiv1\bmod p$,
contradictory to $\gamma\not\equiv 1\bmod p^i$.
  \item[(ii)] $2\cut[i]{\gamma} > p^{i}$.
From $\crr[\gamma]{i}{a_1}=\crr[\gamma]{i}{a_2}$ we get
$\gamma_i\equiv 2\gamma_i + 1 \equiv (\gamma-1)/2\bmod p$,
implying $\gamma_i = p-1$ and $\gamma\equiv -1\bmod p$,
contradictory to $\gamma\not\equiv -1\bmod p^i$.
\end{enumerate}
Therefore,
our assumption $\crr[\gamma]{i}{a_1}=\crr[\gamma]{i}{a_2}=\crr[\gamma]{i}{a_3}$
is ridiculous, and there exist $a,b\in\set{a_1,a_2,a_3}$
satisfying $\crr[\gamma]{i}{a}\neq\crr[\gamma]{i}{b}$.
\fp

\begin{lemma}\label{lemma:new-map-cond-2-3}
A map $\cmap:\pring\rightarrow\GF{p}$ is written as
\begin{equation*}\label{eqn:new-map-str}
\cmap(x)=f_0(x_{e-1})f_1(x_0,x_1,\dots,x_{e-2})
+f_2(x_0,x_1,\dots,x_{e-2}),
\end{equation*}
where $f_0\in\GF{p}[x_{e-1}]$ and
$f_1,f_2\in\GF{p}[x_0,x_1,\dots,x_{e-2}]$.
If $1\leq\deg f_0<p$,
$\left(x_{0}^{p-1}-1\right)\nmid f_1$ and
$x_0\nmid f_1$,
then Statement (ii) and Statement (iii)
of Corollary \ref{thm:branch-p-3} hold.
Particularly, if $f_1=1$, $f_2=0$, and $f_0$
is a permutation polynomial over $\GF{p}$,
then  Statement (ii) and Statement (iii)
of Theorem \ref{thm:uniform-branch-p-3} hold.
\end{lemma}
\pf
Without ambiguity we denote $f_j(x)=f_j(x_0,x_1,\dots,x_{e-2})$ for
$x\in\pring$, $j=1,2$. See $f_j(x+yp^{e-1})=f_j(x)$ for any
$y\in\GF{p}$.

Because $\left(x_{0}^{p-1}-1\right)\nmid f_1$,
we have $(x_0-i)\nmid f_1$ for some $i\in\uGF{p}$,
and hence there exists $a\in\gset$
satisfying $f_1(a)\neq0$. See that $f_2$ is constant on
$a+p^{e-1}\pring$ and $f_0$ is not constant on $a+p^{e-1}\pring$.
Thus, $\cmap$ is not constant on $a+p^{e-1}\pring$ and  Statement
(ii) of Corollary \ref{thm:branch-p-3} holds.

Moreover, because $x_0\nmid f_1$,
there exists $a\in p{\pring}$
satisfying $f_1(a)\neq0$. See that $f_2$ is constant on
$a+p^{e-1}\pring$ and $f_0$ is not constant on $a+p^{e-1}\pring$.
Hence, $\cmap$ is not constant on $a + p^{e-1}\pring$.
Thus,  Statement (iii) of Corollary \ref{thm:branch-p-3} is satisfied.

Now consider the special case:
 $f_1=1$, $f_2=0$, and $f_0$ is a permutation of $\GF{p}$.

Fix any $s\in \GF{p}$, and there there exists $b\in\GF{p}$ 
satisfying $f(b) = s$.
For any $a\in\set{0,1}$, let $c = a + p^{e-1}b$. Then
$\cmap(c)=f(b)=s$ and $\cmap(c+p^{e-1}\pring)=f(\GF{p})=\GF{p}$.
Thus, $\cmap(c+p^{e-1}\pring)\cap D\supset\set{\cmap(a)}$, but
$\cmap$ is not constant on $c +p^{e-1}\pring$.
Take $a=1$ and then Statement (ii) of Theorem
\ref{thm:uniform-branch-p-3} holds.
Take $a=0$ and then Statement (iii) of Theorem
\ref{thm:uniform-branch-p-3} holds.
\fp

Now after preparation, it comes to the proof of Theorem
\ref{thm:uniform-perm-poly}, Theorem \ref{thm:new-map-str},
Theorem \ref{thm:new-map-weak} and
Theorem \ref{thm:new-map-weak-sq}.

\pf[Proof of Theorem \ref{thm:uniform-perm-poly}]
First, we prove
Statement (i) and Statement (ii) of Theorem \ref{thm:uniform-perm-poly}.
By Theorem \ref{thm:uniform-branch-p-3},
it is sufficient to  show that Statements (i)-(iii) of Theorem
\ref{thm:uniform-branch-p-3} hold.
Due to Lemma \ref{lemma:new-map-cond-2-3},
it only remains to prove Statement (i) of Theorem
\ref{thm:uniform-branch-p-3}.

Below we prove Statement (i) of Theorem \ref{thm:uniform-branch-p-3}
respectively for Statement (i) and Statement (ii) of Theorem \ref{thm:uniform-perm-poly}.
Now suppose  $1\neq\omega\in\pring$
and $\omega^r\equiv1\bmod p^e$, where $r>1$ and $r\mid(p-1)$.
\begin{itemize}
  \item Proof of Statement (i) of Theorem \ref{thm:uniform-perm-poly}:
Fix any $s\in D$, and there there exists $a_{e-1}\in\set{0,1,\dots,p-1}$
satisfying $f(a_{e-1}) = s$.
If $\prmpoly$ is strongly primitive, we choose $a = p^{e-1}a_{e-1}$.
Then $\cmap(\omega a) = f(a_{e-1})\neq
f(\omega a_{e-1}) = \cmap(\omega a)$.
Hence, Statement (i) of Theorem \ref{thm:uniform-branch-p-3} holds.
  \item Proof of Statement (ii) of Theorem \ref{thm:uniform-perm-poly}:
Now suppose $\prmpoly$ to be primitive and $f(0)\in D$.
By Lemma \ref{lemma:carry-sum-0}, there
exists $b\in\gset$ satisfying $\crr{e-1}{b}\neq 0$.
Noticing $\crr{e-1}{\cut[e-1]{b}}=\crr{e-1}{b}\neq0$
and $\crd[e-1]{\cut[e-1]{b}}=0$,
we use Lemma \ref{lemma:carry-def-prty} to get
$\cmap(\omega\cut[e-1]{b}) =
f(\crd[e-1]{\omega\cut[e-1]{b}})=
f(\crr{e-1}{\cut[e-1]{b}} + \omega \crd[e-1]{\cut[e-1]{b}})
=f(\crr{e-1}{b})\neq f(0) =\cmap(\cut[e-1]{b})$.
Then $\cmap(\cut{b})\in D$ but $\cmap$ is not constant on $\set{\omega^i \cut[e-1]{b}:1\leq i\leq r}$.
Thus, Statement (i) of Theorem \ref{thm:uniform-branch-p-3} holds.
\end{itemize}
The proof of Statement (i) and Statement (ii) of Theorem \ref{thm:uniform-perm-poly}
is complete.

Now we prove Statement (iii) of Theorem \ref{thm:uniform-perm-poly}.
Clearly, if $\seq{u}=\seq{v}$ then $\indmap{\cmap}(\seq{u})$
and $\indmap{\cmap}(\seq{u})$ are $D$-uniform
for any $\emptyset\neq D\subset\GF{p}$.
It remains to prove that if $\seq{u}\not\equiv\pm\seq{v}\bmod p^e$, then
$\indmap{\cmap}(\seq{u})$ and $\indmap{\cmap}(\seq{u})$ are not
$D$-uniform for any $\emptyset\neq D\subset\GF{p}$.
Suppose $\seq{u}\not\equiv\pm\seq{v}\bmod p^e$.
By Lemma \ref{lemma:map-eqrl},
it is sufficient to prove that
$\cmap$ is not constant on some equivalence class
$C$ w.r.t. $\rlabe$, where $\cmap(C)\cap D\neq\emptyset$.

By Lemma \ref{lemma:equiv-cls-sub},
at least one of the statements (i)-(iii) of Lemma \ref{lemma:equiv-cls-sub}
is true.
\begin{itemize}
  \item
Assume that Statement (i) of Lemma \ref{lemma:equiv-cls-sub}
is true. We have a $k$-th root of unity $\gamma\neq1$,
and $\seq{v}=\gamma\seq{u}$.
By Lemma \ref{lemma:carry-diff},
there exists $a,b\in\set{0,1,\dots,p^{e-1}-1}$ satisfying
$a,b\in\gset$ and $\crr{e-1}{a}\neq\crr{e-1}{b}$.
Fix any $s\in D$ and there exists $a_{e-1}\in\GF{p}$ satisfying
$f(a_{e-1})=s$.
Let $c_1= a + a_{e-1}p^{e-1}$ and $c_2= b + a_{e-1}p^{e-1}$.
By Eq.(\ref{eqn:cmp-map-cry-gnl}),
we notice $\crd[e-1]{a}=\crd[e-1]{b}=0$
and compute
\begin{equation*}
\left\{
  \begin{array}{l}
  \crd[e-1]{\gamma c_1}\equiv \crr{e-1}{a} + \gamma a_{e-1} \bmod p; \\
  \crd[e-1]{\gamma c_2}\equiv \crr{e-1}{b} + \gamma a_{e-1} \bmod p.
  \end{array}
\right.
\end{equation*}
Thus, $\crd[e-1]{\gamma c_1}\neq\crd[e-1]{\gamma c_1}$.
Since $f$ is a permutation of $\GF{p}$,
$\cmap(\gamma c_1) = f(\crd[e-1]{\gamma c_1})\neq
f(\crd[e-1]{\gamma c_2}) = \cmap(\gamma c_2)$.
Seeing $\cmap(c_1)=f(a_{e-1})=\cmap(c_2)$.
We have $\cmap(\gamma c_j)\neq \cmap(c_j)$ for some $j\in\set{1,2}$.
Therefore, for $c_j\in\gset$,
 $\cmap(c_j)=s\in D$ but $\cmap$ is not constant on
the equivalence class
$\set{c_j \gamma^i:1\leq i\leq k}$.
  \item
Assume that Statement (ii) of Lemma \ref{lemma:equiv-cls-sub}
is true.
By Lemma \ref{lemma:new-map-cond-2-3},
Statement (ii) of Theorem \ref{thm:uniform-branch-p-3}
holds.
Similar to the proof of Theorem \ref{thm:uniform-branch-p-3},
in this case $\cmap$ is not constant on some equivalence class
$C$ w.r.t. $\rlabe$, where $\cmap(C)\cap D\neq\emptyset$.
  \item
Assume that Statement (iii) of Lemma \ref{lemma:equiv-cls-sub}
is true.
By Lemma \ref{lemma:new-map-cond-2-3},
Statement (iii) of Theorem \ref{thm:uniform-branch-p-3}
holds.
Similar to the proof of Theorem \ref{thm:uniform-branch-p-3},
in this case $\cmap$ is not constant on some equivalence class
$C$ w.r.t. $\rlabe$, where $\cmap(C)\cap D\neq\emptyset$.
\end{itemize}
Therefore, 
$\cmap$ is not constant on some equivalence class
$C$ w.r.t. $\rlabe$, where $\cmap(C)\cap D\neq\emptyset$,
and Statement (iii) of Theorem \ref{thm:uniform-perm-poly} holds.
\fp

If $\prmpoly$ is not strongly primitive, functions like
Eq.(\ref{eqn:map-zq1}) do not necessarily induce injective maps on
$\gseq$. Below is an example.
\begin{example}
Use the notations and the primitive sequence $\seq{s}$ in Example
\ref{exmp:prim-p-3}. Let $\seq{u} = \seq{s}$ and $\seq{v} =
-\seq{s}$. The map $\cmap: \pring \rightarrow \GF{p}$ is defined as
$\cmap(x)=x_{e-1}^2+x_{e-1}$. Then
$\indmap{\cmap}(\seq{u})=\indmap{\cmap}(\seq{v})$.
\end{example}

In the sequel we give three new families of
entropy-preserving maps from  $\GF{p}^e$ to $\GF{p}$.
\pf[Proof of Theorem \ref{thm:new-map-str}] It is sufficient to show that the
statements (i)-(iii) of Corollary \ref{thm:branch-p-3} hold.

Without ambiguity we denote $f_j(x)=f_j(x_0,x_1,\dots,x_{e-2})$ for
$x\in\pring$, $j=1,2$. See $f_j(x+yp^{e-1})=f_j(x)$ for any
$y\in\GF{p}$.

The condition $f_1(0)\neq0$ implies $x_0\nmid f_1$. Hence,
by Lemma \ref{lemma:new-map-cond-2-3}, Statement (ii)
and Statement (iii) of Corollary \ref{thm:branch-p-3} hold.

Now we use reductio ad absurdum to prove Statement (i) of
Corollary \ref{thm:branch-p-3}.
Suppose that there exists an $r$-th root of unity
$1\neq\omega\in\pring$, where $r$ is a prime divisor of $p-1$, such
that $\cmap(\omega a)=\cmap(a)$ for any $a\in\pring$.

By Corollary \ref{cor:value-occur}, for any $j\in\GF{p}$, $p^{e-1}j$
occurs in any sequence in $\gseq$. Notice that
$\cmap(p^{e-1}x_{e-1})=f_0(x_{e-1})f_1(0)+f_2(0)$. Because
$\cmap(p^{e-1}x_{e-1}\omega)=\cmap(p^{e-1}x_{e-1})$ and
$f_1(0)\neq0$, we have $f_0(x_{e-1})=f_0(\omega x_{e-1})$. Write
$f_0(z) = c_0' + c_1' z + \cdots + c_d' z^{d}$, $1\leq d<p$. Since
$f_0({\omega}z) - f_0(z) = \sum_{i=0}^d c_{i}'({\omega}^i-1)z^i = 0$
is constant over $\GF{p}$, we have $c_{i}' = 0$ for any $r\nmid i$.
Denote $t = \max\set{i\in\Int:i\leq d/r,c_{ri}'\neq0}$ and define a
function on $\GF{p}$ as $g(z) = \sum_{i=0}^{t}c_{ri}' z^i $. We have
$f_0(x_{e-1})=g(x_{e-1}^r)$, where $g\in\GF{p}[x_{e-1}]$ and
$1\leq\deg g< p/r$. See that $\deg g = 0 $ is impossible because
$\deg f_0 = r \deg g \geq 1 $.

By the supposition $\cmap(\omega x)=\cmap(x)$, we have $\cmap(\omega
x + p^{e-1}\omega y)-\cmap(\omega x) = \cmap(x+p^{e-1} y)-\cmap(x) $
for any $y\in\GF{p}$, i.e.,
\begin{equation*}
\left(f_0(\crd[e-1]{\omega (x+p^{e-1}
y)}-f_0(\crd[e-1]{\omega x})\right)f_1(\omega x)
-(f_0(x_{e-1}+y)-f_0(x_{e-1}))f_1(x)=0.
\end{equation*}
By  Eq.(\ref{eqn:cmp-map-cry-gnl}), 
$\crd[e-1]{\omega( x+ y p^{e-1} )} \equiv
\omega (x_{e-1} + y) + \crr{e-1}{x}\bmod p$.
Let  $\Delta = y + {x}_{e-1}$.
Then
\begin{equation*}
f_0(\crd[e-1]{\omega ( x + p^{e-1} y)} =
f_0(\omega \Delta + \crr{e-1}{x}) =
g_0((\omega \Delta + \crr{e-1}{x})^r)=
g_0((\Delta +  \crr{e-1}{x}/\omega)^r),
\end{equation*}
and we have
\begin{equation}\label{eqn:new-map-str-chk}
\left(g\left((\Delta+ \crr{e-1}{x}/{\omega})^r\right)-f_0(\crd[{e-1}]{\omega
x})\right)f_1(\omega x) -
\left(g(\Delta^r)-f_0(x_{e-1})\right)f_1(x)=0,
\end{equation}
for any $\Delta\in\GF{p}$. By Lemma \ref{lemma:new-map-cond-2-3},
there exists $a\in\gset$ with
$f_1(a)\neq0$. Now we consider two cases.
\begin{itemize}
  \item
$f_1(\omega^ia)\neq f_1(a)$ for
some $i\in\set{1,2,\dots,r-1}$.
Let $k=\min\set{1\leq i< r:f_1(\omega^i a)\neq f_1(a)}$
and substitute $\omega^{k-1}a$ for $x$ in Eq.(\ref{eqn:new-map-str-chk}).
Then on the left hand of
Eq.(\ref{eqn:new-map-str-chk}) the term in $\Delta$ of the highest
degree is $c_{rt}'\left(f_1(\omega x)-f_1(x)\right)\Delta^{rt}\neq0$,
which is absurd.

\item  $f_1(\omega^ia)=f_1(a)\neq0$ for
any $i\in\set{1,2,\dots,r}$.
Note that the coefficient of $\Delta^{ct}$ is zero
on the left hand of Eq.(\ref{eqn:new-map-str-chk}).
By Lemma \ref{lemma:carry-sum-0}, there
exists $b\in\set{\omega^ia:i=1,2,\dots,r}$ with $\crr{e-1}{b}\neq0$.
We substitute $b$ for $x$ in Eq.(\ref{eqn:new-map-str-chk}).
Then on the left hand of
Eq.(\ref{eqn:new-map-str-chk}) the term in $\Delta$ of the 
highest degree is $f_1(\omega
b)c_{rt}rt\crr{e-1}{b}\Delta^{rt-1}/\red{\omega}\neq0$,
which is absurd.
\end{itemize}

Therefore, our supposition that $\cmap(\omega a)=\cmap(a)$ for any $a\in\pring$
is ridiculous,
and hence Statement (i) of Corollary
\ref{thm:branch-p-3} is true. \fp

\pf[Proof of Theorem \ref{thm:new-map-weak}] It is sufficient to show
that the statements (i)-(iii) of Corollary \ref{thm:branch-p-3} hold.

We also denote $f_j(x)=f_j(x_0,x_1,\dots,x_{e-2})$ for $x\in\pring$,
$j=1,2$. See $f_j(x+ip^{e-1})=f_j(x)$ for any $i\in\GF{p}$.

By Lemma \ref{lemma:new-map-cond-2-3}, Statement (ii)
and Statement (iii) of Corollary \ref{thm:branch-p-3} hold.

Now we use reductio ad absurdum to prove Statement (i) of
Corollary \ref{thm:branch-p-3}.
Suppose that there exists an $r$-th root of unity
$1\neq\omega\in\pring$, where $r$ is a prime divisor of $p-1$, such
that $\cmap(\omega a)=\cmap(a)$ for any $a\in\gset$.

For any
$x\in\gset$ and any $y\in\GF{p}$,
$\cmap(\omega x+p^{e-1}\omega
y)-\cmap(\omega x)=\cmap(x+p^{e-1}y)-\cmap(x)$,
i.e.,
\begin{equation*}
f_1(\omega x)\left(\crd[e-1]{\omega ( x+p^{e-1} y)}^\ell-\crd[e-1]{\omega
x}^\ell\right) = f_1(x)((x_{e-1}+y)^\ell-x_{e-1}^\ell).
\end{equation*}
By  Eq.(\ref{eqn:cmp-map-cry-gnl}), 
$\crd[e-1]{\omega( x+ y p^{e-1} )} \equiv
\omega (x_{e-1} + y) + \crr{e-1}{x}\bmod p$.
Let  $\Delta =  {x}_{e-1} + y$.
Then
\begin{equation}\label{eqn:new-map-weak-chk}
f_1(\omega x)\left(\left({\omega}\Delta+\crr{e-1}{x}\right)^\ell
-\crd[e-1]{\omega x}^\ell\right) -
f_1(x)\left(\Delta^\ell-x_{e-1}^\ell\right) = 0,
\end{equation}
for any $\Delta\in\GF{p}$.
Computing terms in $\Delta$ of the
(second) highest degree in Eq.(\ref{eqn:new-map-weak-chk}), we have
$f_1(x)={\omega}^\ell f_1(\omega x)$ and $ \crr{e-1}{x}f_1(\omega x)=0$
for $x\in\gset$.
By Lemma \ref{lemma:carry-sum-0}, for any $a\in\gset$, there exists
$k\in\set{1,2,\dots,r}$ satisfying $\crr{e-1}{\omega^ka}\neq0$.
Substituting $\omega^ka$ for $x$ in $ \crr{e-1}{x}f_1(\omega x)=0$, we get
$f_1(\omega^{k+1}a)=0$. Then iteratively substituting $\omega^ja$
for $x$ in $f_1(x)={\omega}^\ell f_1(\omega x)$, $0\leq j \leq i$,
we get $f_1(a)={\omega}^{\ell(k+1)}f_1(\omega^{k+1}a)=0$. Thus,
$f_1(a) = 0$ for any $a\in\gset$, contradictory to
$(x_0^{p-1}-1)\nmid f_1$.

Therefore, our supposition that $\cmap(\omega a)=\cmap(a)$ for any $a\in\gset$
is absurd, and hence Statement (i) of Corollary
\ref{thm:branch-p-3} is true. \fp

\pf[Proof of Theorem \ref{thm:new-map-weak-sq}] It is sufficient to show
that the statements (i)-(iii) of Corollary \ref{thm:branch-p-3} hold.

We denote $f_1(x)=g_0(x_k)+g_1(x_0,x_1,\dots,x_{k-1})$
and  $f_2(x)=f_2(x_0,x_1,\dots,x_{e-2})$ for $x\in\pring$.
See $f_j(x+ip^{e-1})=f_j(x)$ for any $i\in\GF{p}$.

When $k=0$, $f_1(x) = g_0(x_0)$ and
we are given $(x_0^{p-1}-1)\nmid f_1 $ and $x_0\nmid f_1$.
When $1\leq k\leq e-2$, $f_1(x) = g_0(x_k)+g_1(x_0,x_1,\dots,x_{k-1})$
satisfies $(x_0^{p-1}-1)\nmid f_1 $ and $x_0\nmid f_1$.
Thus, by Lemma \ref{lemma:new-map-cond-2-3}, Statement (ii)
and Statement (iii) of Corollary \ref{thm:branch-p-3} hold.

Now we use reductio ad absurdum to prove Statement (i) of
Corollary \ref{thm:branch-p-3}.
Suppose that there exists an $r$-th root of unity
$1\neq\omega\in\pring$, where $r$ is a prime divisor of $p-1$, such
that $\cmap(\omega a)=\cmap(a)$ for any $a\in\gset$.

For any $x\in\gset$ and any $y\in\GF{p}$,
$\cmap(\omega x+p^{e-1}\omega
y)-\cmap(\omega x)=\cmap(x+p^{e-1}y)-\cmap(x)$,
i.e.,
\begin{equation*}
f_1(\omega x)\left(\crd[e-1]{\omega ( x+p^{e-1} y)}-\crd[e-1]{\omega
x}\right) = f_1(x)((x_{e-1}+y)-x_{e-1}).
\end{equation*}
Since $\crd[e-1]{\omega( x+ y p^{e-1} )} \equiv \crd[e-1]{\omega x} + \omega y \bmod p$,
we have $\omega f_1(\omega x)y = f_1(x)y$
for any $y\in\GF{p}$, and conclude that
$\omega f_1(\omega x) = f_1(x)$ for $x\in\gset$.

If $k=\deg g_0 =0$, then $f_1\in\uGF{p}$ is constant, contradictory
to $f_1(x)={\omega} f_1(\omega x)$. Here $f_1=g_0=0$ is impossible
because $x_0\nmid g_0$.

Otherwise, consider $k\geq1$ or $\deg g_0>k=0$.
For any $x\in\gset$
and $\Delta\in\GF{p}$,
$ {\omega}(f_1(\omega(x+p^{k}\Delta))-f_1(\omega x))=
f_1(x+p^{k}\Delta)-f_1(x)$, i.e.,
\begin{equation}\label{eqn:new-map-weak-chk-2}
   {\omega}\left(
    g_0({\omega}\Delta + \crd[k]{\omega x}  ))-g_0(\crd[k]{{\omega} x})
    \right) = g_0(x_k + \Delta) - g_0(x_k).
\end{equation}
Comparing the terms in $\Delta$ of the highest degree in
Eq.(\ref{eqn:new-map-weak-chk-2}), we have ${\omega}^{1+\deg
g_0}\equiv1\bmod p$
 and conclude $r\mid (\deg g_0+1)$.
However, if $\gcd(p-1,\deg g_0+1)=1$, then $r\nmid (\deg g_0+1)$
since $r\mid(p-1)$. Here comes a contradiction.

Therefore, our supposition that
$\cmap(\omega a)=\cmap(a)$ for any $a\in\gset$
is absurd and Statement (i) of Corollary
\ref{thm:branch-p-3} is true. \fp


Finally, since modular functions are a kind of important
compressing maps,
we also give another proof of entropy-preservation of the
modular compression \cite{ZQ08}.
\begin{theorem}[Zhu-Qi]\label{lemma:mod-func}
Let $\cmap(x)=\rep{x}\bmod M$ be a map defined on $\pring$, where
the positive integer $M\geq2$ is not a power of $p$. If $\prmpoly$
is primitive, then $\indmap{\cmap}$ is injective on $\gseq$.
\end{theorem}
\pf It is sufficient to show that the statements (i)-(iii)
of Corollary \ref{thm:branch-p-3} hold.

Suppose $1<\omega<p^e$ and $\omega^r\equiv1\bmod p^e$, where $r>1$
and $r\mid (p-1)$. Since $\omega\not\equiv1\bmod p$, there exists $a\in\gset$ satisfying $a\equiv
1/(\omega-1)\mod p^e$. Clearly, $a<p^e-1$. Hence,
$\rep{a\omega}\equiv a+1\bmod p^e$ and we have
$\cmap(a\omega)\neq\cmap(a)$. Thus, Statement (i) of Corollary
\ref{thm:branch-p-3} holds.

See that $M$ is not a power of $p$. Hence, for any $a\in\pring$ we
have
$0\leq \cut[e]{a+jp^{e-1}} <
\cut[e-1]{a+(j+1)p^{e-1}}<p^e$
for some
$j\in\set{0,1,\dots,p-1}$.
Then $\cut[e-1]{a+(j+1)p^{e-1}} - \cut[e]{a+jp^{e-1}} = p^{e-1}$
and hence
$\cmap\left(a+jp^{e-1}\right)\neq\cmap\left(a+(j+1)p^{e-1}\right)$.
Thus, Statement (ii) and Statement (iii) of Corollary
\ref{thm:branch-p-3} are true. \fp

\subsection{The number of conditioned primitive polynomials
}\label{subsect:no-prim-seq}
In this subsection
we give the number of primitive polynomials,
the number of  strongly primitive polynomials, and
the number of  primitive polynomials
satisfying Condition \ref{cond:pit-p-lt-3}.

We characterize such conditioned polynomials
in Lemma \ref{lemma:unique-factor}
and Lemma \ref{lemma:poly-determine},
and then count them in Theorem \ref{fact:count-poly}.

\begin{lemma}\label{lemma:unique-factor}
Let $f,g,f',g'$ be monic polynomials over $\pring$ satisfying
$fg=f'g'$, $f\equiv f'\bmod p$ and $g\equiv g'\bmod p$. If $f\bmod
p$ and $g\bmod p$ are relatively prime over $\GF{p}$, then $f=f'$
and $g=g'$.
\end{lemma}
\pf Denote  $\Delta_f = f' - f $ and $\Delta_g = g'- g$. Because the
monic polynomials $f,g,f',g'$ satisfy
 $f\equiv f'\bmod p$ and $g\equiv g'\bmod p$,
we have
\begin{equation}\label{eqn:uni-factor}
    \left\{
     \begin{array}{l}
        \deg \Delta_f < \deg f,\\
        \deg \Delta_g < \deg g,\\
        \Delta_f \equiv \Delta_g\equiv 0\bmod p.
     \end{array}\right.
\end{equation}
Let $k=\max\set{1\leq i\leq e:p^i\mid \Delta_f, p^i\mid \Delta_g}$.
Suppose $k<e$, i.e. $\Delta_f=\Delta_g=0$ does not hold over
$\pring$. Since
\begin{equation*}
 f'g' = (f+\Delta_f)(g+\Delta_g) =
fg + g\Delta_f +f\Delta_g +\Delta_f\Delta_g = fg ,
\end{equation*}
we have $g\Delta_f + f\Delta_g\equiv0\bmod p^{k+1}$, i.e.,
\begin{equation*}
  g\cdot(\Delta_f/p^k) +f\cdot(\Delta_g/p^k)\equiv0\bmod p,
\end{equation*}
implying
\begin{equation*}
\left\{
\begin{array}{ccc}
(f\bmod p)&\mid& (g(\Delta_f/p^k)\bmod p),\\
(g\bmod p)&\mid& (f(\Delta_g/p^k)\bmod p).
\end{array}\right.
\end{equation*}
Furthermore, since $f\bmod p$ and $g\bmod p$ are relatively prime,
we have
\begin{equation*}
\left\{
\begin{array}{ccc}
(f \bmod p)&\mid &((\Delta_f/p^k)\bmod p),\\
(g \bmod p)&\mid &((\Delta_g/p^k)\bmod p).
\end{array}\right.
\end{equation*}
Besides, by Eq.(\ref{eqn:uni-factor}),
 $\deg \Delta_f < \deg f$ and $\deg \Delta_g < \deg g$.
Thus, we have $(\Delta_f/p^k)\equiv(\Delta_g/p^k)\equiv0\bmod p$, i.e.,
$p^{k+1}\mid \Delta_f$ and $p^{k+1}\mid \Delta_g$, contradictory to
the definition of $k$ above. Thus, our supposition $k<e$ is absurd
and hence $\Delta_f=\Delta_g=0$, i.e. $f=f'$ and $g=g'$. \fp

\begin{lemma}\label{lemma:poly-determine}
Let $f,g$ be two monic polynomials of degree $\len$ over $\pring$.
Assume that $f\bmod p$ and  $g\bmod p$ are irreducible over
$\GF{p}$. Let $h_f$ (resp. $h_g$) be a polynomial of degree less than
$\len$ satisfying $x^{p^\len-1} - 1 \equiv ph_f\bmod f$ (resp.
$x^{p^\len-1} - 1 \equiv ph_g\bmod g$). If $f\equiv g\bmod p$ and
$h_f\equiv h_g\bmod p^{e-1}$, then $f=g$.
\end{lemma}
\pf There exist polynomials $\lambda,\lambda'$ satisfying
\begin{equation*}
    \left\{
     \begin{array}{ccc}
x^{p^\len-1} - 1 &=& ph_f + \lambda f,\\
x^{p^\len-1} - 1 &=& ph_g + \lambda'g.
     \end{array}\right.
\end{equation*}
First, we have $\lambda f=\lambda'g$ because $h_f\equiv h_g\bmod
p^{e-1}$. Second, $\lambda f\equiv x^{p^\len-1} - 1\equiv\lambda'
g\bmod p$. Since $\left(x^{p^\len-1} - 1\right)\bmod p$ has no
multiple root over $\GF{p}$ and $f\equiv g\bmod p$, we also have
$\lambda\equiv \lambda'\bmod p$ and conclude that $f\bmod p$ and
$\lambda\bmod p$ are relatively prime over $\GF{p}$. Then by Lemma
\ref{lemma:unique-factor}, $f=g$. \fp

\begin{theorem}\label{fact:count-poly}
Let $\phi$ denote the Euler totient function. Then the number of
primitive polynomials of degree $\len$ is
$p^{\len(e-2)}(p^\len-1)\phi(p^\len-1)/\len$;
the number of strongly primitive polynomials of degree $\len$ is
$p^{\len(e-2)}(p^\len-p)\phi(p^\len-1)/\len$;
and the number of primitive polynomials of degree $\len$ satisfying
$\left((x^{p^\len-1}-1)^2/p^2\bmod (p,\prmpoly)\right) \notin\GF{p}$
is
\begin{equation*}
   p^{\len(e-2)}\left(p^\len-p-(p-1)\left(1+(-1)^\len\right)/2\right)\phi(p^\len-1)/\len.
\end{equation*}
\end{theorem}
\pf By Lemma \ref{lemma:poly-determine}, a primitive polynomial
$\prmpoly$ is uniquely determined by the primitive polynomial
$\prmpoly \bmod p$ and the polynomial $\delta$ over
$\pring[p^{e-1}]$ of degree less than $\len$.

Above all, it is known that there are $\phi(p^\len-1)/\len$
primitive polynomials over $\GF{p}$ of degree $\len$.

To ensure $\prmpoly$ to be primitive over $\pring$, we only have to
choose $\delta$ over $\pring[p^{e-1}]$ of degree less than $\len$
satisfying $\delta\not\equiv0\bmod p$. Since
\begin{align*}
&\cset{\set{g\in\pring[p^{e-1}][x]:\deg g<\len,g\not\equiv0\bmod p}}\\
=&\cset{\set{g\in\pring[p^{e-1}][x]:\deg g<\len}}-
\cset{\set{g\in\pring[p^{e-1}][x]:\deg g<\len,g\equiv0\bmod p}}\\
=& p^{(e-1)\len} - p^{(e-2)\len} = p^{\len(e-2)}(p^\len-1),
\end{align*}
the number of primitive polynomials of degree $\len$ is
$p^{\len(e-2)}(p^\len-1)\phi(p^\len-1)/\len$.

To ensure $\prmpoly$ to be strongly primitive over $\pring$, we only
have to choose $\delta$ over $\pring[p^{e-1}]$ of degree less than
$\len$ satisfying $(\delta\bmod p)\notin\GF{p}$. Since
\begin{align*}
&\cset{\set{g\in\pring[p^{e-1}][x]:\deg g<\len,\forall a\in\GF{p}, g\not\equiv a\bmod p}}\\
=&\cset{\set{g\in\pring[p^{e-1}][x]:\deg g<\len}}\\
&-
\cset{\set{g\in\pring[p^{e-1}][x]:\deg g<\len,\exists a\in\GF{p}, g\equiv a\bmod p}}\\
=& p^{(e-1)\len} - p\cdot p^{(e-2)\len} = p^{\len(e-2)}(p^\len-p),
\end{align*}
the number of strongly primitive polynomials of degree $\len$ is
$p^{\len(e-2)}(p^\len-p)\phi(p^\len-1)/\len$.

To ensure $\prmpoly$ to satisfy $\left(\delta^2\bmod
(p,\prmpoly)\right) \notin\GF{p}$ we only have to choose $\delta$
over $\pring[p^{e-1}]$ of degree less than $\len$ satisfying
$\delta^2 \not \equiv a\bmod (p,\prmpoly) $ for any $a\in\GF{p}$.
Notice that $\GF{q}[x]/(\prmpoly\bmod p)$ is a finite field of
$p^\len$ elements and $\delta\bmod p$ can be considered as an
element, where $\deg\delta<\len$. Since the multiplicative group of
$\GF{p}[x]/(\prmpoly\bmod p)$ is
 a cyclic group of order $p^\len-1$, and
$a\in\uGF{p}$ if and only if $a^{p-1}\equiv 1\bmod p$, we get
\allowdisplaybreaks
\begin{align*}
  & \cset{\set{{g}\in\pring[p^{e-1}][x]:
  \deg g<\len, \forall a\in\GF{p},{g}^2\not\equiv a\bmod (p,\prmpoly)}}\\
=&  \cset{\set{{g}\in\pring[p^{e-1}][x]:
  \deg g<\len}} \\
  &-
  \cset{\set{{g}\in\pring[p^{e-1}][x]:
  \deg g<\len, \exists a\in\GF{p},{g}^2\equiv a\bmod (p,\prmpoly)}}\\
= & p^{\len(e-1)} -
  p^{\len(e-2)}\cset{\set{{g}\in\GF{p}[x]:
  \deg g<\len, \exists a\in\GF{p},g^2\equiv a\bmod (\prmpoly\bmod p)}}\\
= & p^{\len(e-1)} -
  p^{\len(e-2)}\left(1+\cset{\set{{g}\in\GF{p}[x]:
  \deg g<\len, \exists a\in\uGF{p},g^2\equiv a\bmod (\prmpoly\bmod p)}}\right)\\
= &  p^{\len(e-1)} -
  p^{\len(e-2)} \left(1+\cset{\set{0\leq i\leq p^\len-2: (p^\len-1)\mid 2i(p-1)}}\right)\\
= &\left\{
\begin{array}{l@{\text{ if }}l}
 p^{\len(e-1)} -
  p^{\len(e-2)}\left(1+\cset{\set{0\leq i\leq p^\len-2: \frac{p^\len-1}{p-1}\mid i}}\right), &  2\nmid\len,  \\
p^{\len(e-1)} -
  p^{\len(e-2)}\left(1+\cset{\set{0\leq i\leq p^\len-2: \frac{p^\len-1}{2(p-1)}\mid i}}\right), &  2\mid \len,  \\
\end{array}\right.\\
= &\left\{
\begin{array}{l@{\text{ if }}l}
 p^{\len(e-1)} -
  p^{\len(e-2)} p = p^{\len(e-2)}(p^\len-p), &  2\nmid\len,  \\
 p^{\len(e-1)} -
  p^{\len(e-2)}(2p-1)=p^{\len(e-2)}(p^\len-2p+1), &  2\mid \len.  \\
\end{array}\right.
\end{align*}
Thus, the number of primitive polynomials of degree $\len$
satisfying $\left(\delta^2\bmod (p,\prmpoly)\right) \notin\GF{p}$
is
$p^{\len(e-2)}(p^\len-p-(p-1)(1+(-1)^\len)/2)\phi(p^\len-1)/\len$.
\fp

\section{Conclusion}\label{sect:conclusion}
The compressing map extracts nonlinear sequences from
primitive sequences over residue rings.
We characterized the compressing maps such that
the distribution of some elements in the compressed sequence
determines a unique original primitive sequence.
For at least $1-2(p-1)/(p^\len-1)$ of primitive polynomials of degree
$\len$, we gave a clear criterion of injectivity w.r.t. $D$-uniformity
and entropy-preservation, and
also estimated the number of entropy-preserving maps.
Furthermore, we also present specific injective maps
w.r.t. $D$-uniformity and
three kinds of entropy-preserving maps from $\pring$ to $\GF{p}$.

Finally, we comment on relevant work.
In this paper we introduce the language of binary relations
to characterize uniformity. Following the same idea,
it is natural to use binary relations to interpret $s$-uniformity with $\seq{\alpha}$
and $s$-uniformity with $\seq{\alpha}|_k$.
Recall $\seq{\alpha}=\delta\seq{u}\bmod p$.
To tell whether $\seq{u}$ and $\seq{v}$
are $s$-uniform with $\seq{\alpha}$, the key
is to give the equivalence closure of the relation $\brlab$;
To tell whether $\seq{u}$ and $\seq{v}$
are $s$-uniform with $\seq{\alpha}|_k$, the key
is to give the equivalence closure of the relation
\begin{equation*}
 \{(a,b)\in\pring\times\pring:\exists t\in \Int,
    \delta\seq{u}(t)\equiv k\bmod p,
    \seq{u}(t)\equiv a\bmod p^e,
    \seq{v}(t)\equiv b\bmod p^e
    \}.
\end{equation*}
Besides, it is also natural to ask whether
it is possible to transplant the methods of this paper
and its results to the case $p=2$.

\section*{Acknowledgment}
The authors would like to thank the anonymous referees for their
valuable suggestions which help to improve the manuscript.
{L. Wang is supported by the Applied Basic Research Program
of the Sichuan Province, P.~R.~China (2011JY0143),
by CETC Innovation Foundation (JJ-QN-2013-33) and
also partially by Natural Science Foundation of China (61309034).}
{Z. Hu is supported by National Natural Science Foundation of China (Grant No. 61272499)}.

\end{document}